\documentclass[journal]{IEEEtran}

\usepackage{amsfonts} 
\usepackage[dvipsnames]{xcolor}
\usepackage[utf8]{inputenc}
\usepackage[english]{babel}
\usepackage{multirow}
\usepackage{booktabs}
\newcommand{\rev}[1]{\textcolor{black}{#1}}
\newcommand{\revv}[1]{\textcolor{black}{#1}}

\usepackage{cite}

\ifCLASSINFOpdf
   \usepackage[pdftex]{graphicx}
\else
\fi

\usepackage{mathtools}
\usepackage{amsthm}
\newtheorem{lem}{Lemma}
\usepackage{algorithmic}
\usepackage[caption=false,font=footnotesize,labelfont=rm,textfont=rm]{subfig}
\usepackage{dblfloatfix}
\usepackage[hyphens]{url}
\begin{document}
\title{ECO: Enabling Energy-Neutral IoT Devices through Runtime Allocation of Harvested Energy}

\author{Yigit~Tuncel,
        Ganapati~Bhat,
        Jaehyun~Park,
        and~Umit~Ogras%
\thanks{Y. Tuncel and U. Ogras are with the Department
of Electrical and Computer Engineering, University of Wisconsin-Madison, Wisconsin,
WI, 53706 USA. e-mail: \{tuncel, uogras\}@wisc.edu}%
\thanks{G. Bhat is with the School of Electrical Engineering and Computer Science, Washington State University, Pullman, WA, 99164, USA. e-mail: ganapati.bhat@wsu.edu}%
\thanks{J. Park is with the Department of Electrical, Electronic and Computer Engineering, University of Ulsan, 93 Daehak-ro, Mugeo-dong, Nam-gu, Ulsan, Republic of Korea. e-mail: jaehyun@ulsan.ac.kr}%
\thanks{An earlier version of this paper was presented at the 2017 IEEE/ACM International Conference on Computer-Aided Design (ICCAD)~\cite{bhat2017near} and was published in its proceedings (doi: 10.1109/ICCAD.2017.8203801). }%
}

\markboth{}%
{Tuncel \MakeLowercase{\textit{et al.}}: ECO: Enabling Energy-Neutral IoT Devices}

\maketitle

\begin{abstract}
Energy harvesting offers an attractive and promising mechanism to power low-energy devices. However, it alone is insufficient to enable an energy-neutral operation, which can eliminate tedious battery charging and replacement requirements. Achieving an energy-neutral operation is challenging since the uncertainties in harvested energy 
undermine the quality of service requirements. 
\rev{To address this challenge, we present a runtime energy-allocation framework that optimizes the utility of the target device under energy constraints using a rollout algorithm, which is a sequential approach to solve dynamic optimization problems.}
The proposed framework uses an efficient iterative algorithm to compute initial energy allocations at the beginning of a day.
The initial allocations are then corrected at every interval to compensate for the deviations from the expected energy harvesting pattern.
We evaluate this framework using solar and motion energy harvesting modalities and \textit{American Time Use Survey} data from 4772 different users. 
{Compared to prior techniques, the proposed framework achieves up to 35\% higher utility even under energy-limited scenarios.}
\rev{Moreover, measurements on a wearable device prototype show that the proposed framework has 1000x smaller energy overhead than iterative approaches with a negligible loss in utility.}
\end{abstract}


\IEEEpeerreviewmaketitle

\newcommand\EBvec{
\begin{bsmallmatrix}
    E_1^B & E_2^B & \cdots  & E_T^B
\end{bsmallmatrix}^T}

\newcommand\EB{
\begin{bmatrix}
    E_1^B  \\
    E_2^B  \\
    \vdots  \\
    E_T^B
\end{bmatrix}}

\newcommand\ECvec{
\begin{bsmallmatrix}
    E_0^A & E_1^A & \cdots & E_{T-1}^A
\end{bsmallmatrix}^T}

\newcommand\EC{
\begin{bmatrix}
    E_0^A  \\
    E_1^A  \\
    \vdots  \\
    E_{T-1}^A
\end{bmatrix}}

\newcommand\bvec{
\begin{bsmallmatrix}
    E_{min}  & \cdots & E_{min}  &E_{target}
\end{bsmallmatrix}^T}

\newcommand\mvec{
\begin{bsmallmatrix}
    E_{max}  & \cdots & E_{max}  &E_{max}
\end{bsmallmatrix}^T}
 
\newcommand\hvec{
\begin{bsmallmatrix}
    E_0^B + E_0^H & E_1^H & E_2^H \cdots  & E_{T-1}^H
\end{bsmallmatrix}^T}

\newcommand\uvec{
\begin{bmatrix}
    \beta^0 u(E^A_0)  & \beta^1 u(E^A_1) & \cdots & \beta^{T-1} u(E^A_{T-1})
\end{bmatrix}^T}

\newcommand\graduvec{
\begin{bmatrix}
    \beta^0 \frac{\partial u(E^A_0)}{\partial E^A_0}  & \beta^1 \frac{\partial u(E^A_1)}{\partial E^A_1} & \cdots & \beta^{T-1} \frac{\partial u(E^A_{T-1})}{\partial E^A_{T-1}}
\end{bmatrix}}

\newcommand\Lmat{
\begin{bsmallmatrix}
    1 & 0 & 0 & \cdots & 0 \\
    1 & 1 & 0 & \cdots & 0 \\
    1 & 1 & 1 & \cdots & 0 \\
    \vdots & \vdots & \vdots & \ddots & \vdots \\
    1 & 1 & 1 & \cdots & 1
\end{bsmallmatrix}}

\section{Introduction}
\label{sec:intro}

\IEEEPARstart{I}{nternet} of things (IoT) and wearable devices have prevailed in recent years. Their number is expected to grow to 30 billion by 2025 with the ever-reducing power consumption of technology nodes and with new and emerging applications~\cite{iotanalytics,techjury,lau2019survey}.
Battery charging and replacement require significant manual effort and environmental waste.
Furthermore, recent studies show that frequent recharging requirements hinder the adoption of wearable IoT devices in healthcare~\cite{ozanne2018wearables, johansson2018wearable}.
Therefore, harvesting ambient energy has become an attractive solution to power low-power IoT and wearable devices.

Energy harvesting (EH) is the ability to transform ambient energy into usable electrical energy~\cite{paradiso2005energy}.
EH solutions ideally aim for \emph{energy neutrality}, which ensures that the energy consumption in a given period, such as a day, is equal to the energy harvested during the same period.
\rev{Hence, energy-neutral operation can automate battery charging through harvested energy such that the battery level is restored at the end of each day. 
Consequently, the battery can be used without manual charging throughout its lifetime.}
Achieving energy-neutral operation is challenging due to the conflict between the uncertainty in harvested energy and quality of service (QoS) requirements of the target device. 
For example, consider a wearable health application where the target device must collect the vital signals and process them locally to detect abnormalities. 
On the one hand, the device needs a steady and sufficient amount of energy to perform its intended operation, e.g., analyzing the collected signals within a specified amount of time. 
On the other hand, the harvested energy may fluctuate widely or even vanish entirely during the same period. 
\rev{State-of-the-art research and commercial products on energy harvesting for IoT typically tackle such issues by improving the harvested power and reducing the consumed power~\cite{cordis, energiot, atmosic, onio}. 
For example, Energiot designs energy harvesters tailored for specific use-cases to increase the harvested energy~\cite{energiot}. 
Similarly, Atmosic designs low-power wireless technologies tailored for energy harvesting devices to enable energy-neutral operation~\cite{atmosic}. 
Even though those aspects are critical, the very limited and highly varying nature of the harvested energy necessitates deliberate usage of the available energy. 
In this sense, our goal is to maximize the utilization of the target device (i.e., the amount of time QoS requirements are satisfied) under dynamic energy consumption requirements that guarantee energy neutrality.}

\begin{figure*}[!t]
	\centering
	\includegraphics[width=0.95\linewidth]{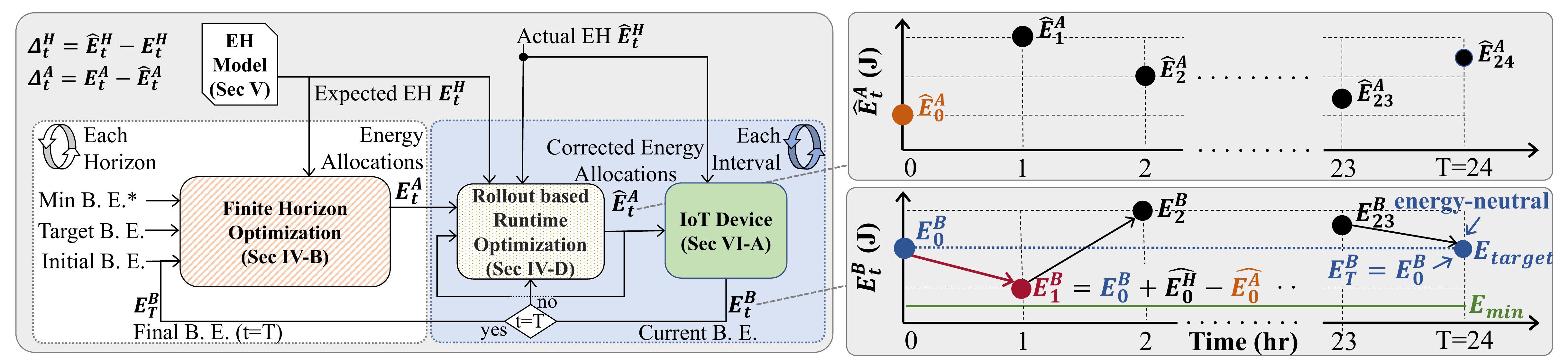}
	\vspace{-4mm}
	\caption{The proposed energy allocation framework (ECO). *B.E.: Battery Energy}
	\vspace{-3mm}
	\label{fig:block}
\end{figure*}

This paper proposes a runtime framework to 
enable \underline{e}nergy-neutrality through \underline{c}onstrained \underline{o}ptimization (ECO) of the harvested energy optimally over a finite horizon divided into uniform time intervals.
\rev{The main motivation behind ECO is to make judicious use of harvested energy so that the application utility is maximized. 
It is crucial to carefully allocate the harvested energy throughout the day because the device does not have continuous access to ambient energy. 
For example, light energy is available only when the user is outdoors during the day or is in a room with sufficient light. 
At the same time, the device needs energy to perform the application tasks throughout the day. 
Therefore, a portion of the harvested energy must be stored in the battery and used when there is no ambient energy available.
}
ECO computes the amount of energy that the device can consume in each time interval to maximize its utility and guarantee an energy-neutral operation.
The fundamental components of ECO are illustrated in Figure~\ref{fig:block}.
It takes the initial battery energy at the beginning of each day and the expected amount of harvested energy during each time interval of the finite horizon (e.g., 24-hour) as inputs.
We obtain the expected amount of harvested energy from a novel EH model that combines light EH and piezoelectric motion EH modalities, as described in Section~\ref{sec:EHModel}.
\rev{However, the operation of the ECO framework is not limited to this particular model. 
It is compatible with any other energy harvesting model.}
We also specify two constraints on the battery energy: i) the minimum battery energy level allowed at any point in time, ii) the battery energy target at the end of the day. 
\rev{With the target energy constraint, the day to day battery operations of recharging and discharging are hidden from the user. Therefore, the users do not need to worry about manually recharging the battery.}

The proposed framework maximizes a utility function under the battery energy constraints.
The utility function can be constructed to model design metrics, such as device throughput and classification accuracy. 
Unlike prior work that requires a logarithmic utility function~\cite{bhat2017near}, ECO can handle an arbitrary function, which can be constructed by measured data.
It first finds the energy that the device can consume during each time interval using the harvested energy profile. For example, suppose that the finite horizon is one whole day divided into 24 one-hour intervals. 
At the beginning of each day, ECO allocates the expected amount of harvested energy during a day into one-hour intervals to maximize the utility function under the battery energy constraints. 
This solution is not optimum due to the variations in harvested energy.
Therefore, the initial allocations are corrected after each time interval using the actual harvested energy.
To find the optimal amount of corrections for the remaining intervals, we propose a lightweight runtime optimization algorithm. 
As a result, the proposed framework adapts to the deviations from the expected EH values with negligible runtime overhead. 

We demonstrate the proposed framework using a low-power wearable IoT device and a dataset that consists of expected and actual energy harvesting values of 4772 users. 
To establish the optimal reference, our results are compared against an iterative gradient projection algorithm. 
{ECO achieves an average utility within 2\% of the iterative gradient projection algorithm with 1000$\times$ smaller overhead.
When compared to prior approaches in literature~\cite{kansal2007power,bhat2017near}, it achieves up to 35\% higher average utility while maintaining negligible runtime overhead.} 

\vspace{1mm}
\textit{The major contributions of this work are as follows:}
\begin{itemize}
\item We present an energy allocation framework (ECO) that enables \emph{energy-neutral operation} under varying EH conditions 
while maximizing the device utility, 

\item ECO supports arbitrary utility functions extracted from real applications,

\item To make realistic evaluations, we construct a novel dataset using light/motion EH modalities and the American Time Use Survey (ATUS)~\cite{amtus} of 4772 users, 

\item ECO achieves up to 35\% higher utility than prior runtime techniques and only 2\% lower utility than an optimal iterative algorithm with negligible runtime overhead.

\end{itemize}

The rest of the paper is organized as follows. We review the related work in Section~\ref{sec:relwork}.
We present an overview of the problem in Section~\ref{sec:overview}.
Section~\ref{sec:optem} gives detailed explanations and mathematical formulations for each of the components in Figure~\ref{fig:block}.
In Section~\ref{sec:EHModel}, we explain the steps we took to pre-process the ATUS dataset and obtain the EH model.
Finally, we evaluate and discuss the results in Section~\ref{sec:eval} and conclude the paper in Section~\ref{sec:conc}.
\section{Related Work}
\label{sec:relwork}

Research and development of the \rev{Internet of Things} have grown considerably in the recent past~\cite{atzori2010internet}. This growth has been enabled by advances in sensor technologies, low-power processing, and communication technologies. IoT devices have applications in a wide range of fields such as wearable health monitoring, environmental monitoring, smart cities, digital agriculture, and robotics~\cite{sutton2019blitz,abdelgawad2016structural,hassanalieragh2015health,jayaraman2015addressing,bhat2020self,lau2017sensor}. 
IoT devices are deployed either as a network of multiple devices or single nodes, depending on the application requirements. For instance, environmental monitoring and smart city applications typically use a wide-area network along with infrastructure to cover the area of interest~\cite{sutton2019blitz,jayaraman2015addressing,sun2020large,zhang2020energy}. Similarly, health monitoring applications use a single node or smaller body area networks~\cite{wan2018wearable,khan2016flexible}.

One of the challenges of IoT devices is limited battery size due to small form-factor, especially in wearable devices. 
The small battery severely constrains the energy budget of the device and battery lifetime. 
Recent work has proposed harvesting energy from ambient sources, such as light, motion, electromagnetic waves, and body heat, to alleviate the issue.
\rev{Prior studies have shown that ambient light has the highest potential for EH with power levels of $>$1 mW and $<$100 $\mu$W in outdoor (5000 lux) and indoor (500 lux) conditions, respectively, with an 8.1 cm$^2$ flexible PV-cell~\cite{jokic2017powering}.
Similarly, human motion EH can yield about 15 $\mu$W with a 23.8 cm$^2$ piezoelectric patch during walking~\cite{tuncel2020towards}. 
In comparison, RF EH harvests 10 $\mu$W with an 18.4 cm$^2$ flexible antenna with -10 dBm @ 915 MHz~\cite{nguyen2018hybrid}, and body heat EH has power levels of about 3 $\mu$W with a 1 cm$^2$ flexible harvester at 15 $^{\circ}$C ambient temperature (i.e., temperature difference of 22 $^{\circ}$C)~\cite{huu2018flexible}.}

Energy harvesting requires algorithms that allocate the harvested energy to the device so that sufficient energy is available at all times of the day~\cite{kansal2007power,vigorito2007adaptive}.
For instance, light energy is typically available during the day. 
Thus, the device has to store sufficient energy for use during the night.
Kansal et al.~\cite{kansal2007power} present the general framework of energy-neutral operation for energy harvesting devices. Energy-neutral operation (ENO) ensures that the total energy consumed in a given period, such as a day, is equal to the energy harvested during the same period.
To achieve this, the authors propose a linear programming approach that adapts the device's duty cycle under energy constraints.
The work in~\cite{buchli2014dynamic} proposes a long-term energy management algorithm, referred to as long term-ENO, that aims to achieve energy neutrality for one year or more.
Similarly, Geissdoerfer et al.~\cite{geissdoerfer2019getting} propose a feedback controller to achieve long-term-ENO.
While these approaches are helpful, they do not consider the application requirements, user activity patterns, or the stochastic nature of energy harvesting.

The stochastic nature of energy harvesting has prompted researchers to investigate approaches that aim to account for the stochasticity~\cite{pughat2015review,dang2015unified,ahmed2017optimal,ku2015data}. These approaches typically model stochastic optimization as a Markov decision process and develop policies to perform the energy management. Unfortunately, these approaches still do not consider the stochasticity arising from user activities.
For example, the harvested energy changes significantly if the user moves indoors after being in the sun. User and context-aware approaches aim to address this issue by leveraging knowledge about the user's activities to reduce the power consumption of the device~\cite{starliper2019activity,alhassoun2019context}. 
These approaches focus only on power management and do not account for energy harvesting. 
Therefore, there is a need for approaches that account for variations in harvested energy due to environmental conditions and user activities.

This work presents a novel approach to enable energy-neutral operation under varying energy harvesting conditions.
Our prior work addressed a similar problem~\cite{bhat2017near}. 
However, it relaxes the minimum battery energy constraints and assumes an impractical logarithmic utility function to derive a closed-form solution. 
In contrast, the ECO framework uses a light-weight rollout technique that supports arbitrary utility functions and finds the energy allocations without relaxing any constraints. 
The rollout phase implicitly accounts for the harvested energy variations due to environmental conditions and user activities. Specifically, we start with an initial energy allocation using the expected EH values.
Then, we use the rollout algorithm at runtime to make adjustments to the energy allocation as a function of the deviations in energy harvesting.
Furthermore, the proposed technique is evaluated using ATUS data of 4772 users with light and motion EH modalities.

\section{Overview}
\label{sec:overview}
This section introduces the battery energy dynamics and the corresponding optimization constraints.
It also presents the notion of the utility function for an arbitrary application.

\vspace{1mm}
\noindent \textbf{Battery energy dynamics:}
The proposed allocation algorithm determines the amount of energy that can be consumed in a given interval based on the current battery level and harvested energy.
Let $T$ denote the finite time horizon of interest. We divide the time horizon into uniform time intervals $t$. 
For example, if we consider a one-day horizon divided into 24 one-hour intervals, $T=24$ hours and $t \in \mathcal{S_T} = {[0,1), [1,2), \ldots, [23,24)}$, as illustrated in Figure~\ref{fig:block}. 
Using this notation, we denote the battery energy at the beginning of interval $[t,t+1)$ as $E^B_t$.
The battery energy at the end of interval $t \in \mathcal{S_T}$ depends on three factors as follows:
\begin{enumerate}
    \item The starting battery energy $E^B_{t}$.
    \item The harvested energy during the current interval $\widehat E^H_t$.
    \item The allocated energy during the current interval $\widehat E^A_t$.
\end{enumerate}
The battery energy level at the end of interval $t$ equals 
to the level at the beginning of interval $t+1$.
Similarly, the battery energy level at the end of finite horizon 
is carried to the next finite horizon (e.g. next day). 
Hence, the battery energy dynamics can be expressed as
\begin{equation}\label{eqn:battery_energy}
E^B_{t+1} = E^B_t + \eta \widehat E^H_t - \widehat E^A_t, \hspace{1mm} t \in \mathcal{S_T}
\hspace{2mm} \mathrm{and} \hspace{2mm} E^B_{T} = E^B_0
\end{equation}

\rev{Here, $\eta$ accounts for the inefficiencies at the energy harvesting and battery charging interfaces.} The battery level must be bounded from below to ensure that the target device has sufficient reserves ($E_{min}$) to execute an emergency task. 
\rev{In addition, battery level can not exceed the the battery capacity.
To achieve these objectives, we introduce the following minimum and maximum energy constraints:
\begin{equation}\label{eqn:minEnergy}
E_{max} \geq  E^B_{t} \geq  E_{min} \hspace{2mm}   t \in \mathcal{S_T}
\end{equation}
}
To achieve energy-neutral operation, we also need to ensure that the battery energy level in the beginning of each finite horizon is greater than or equal to a specified target. 
For example, if the battery energy level in the beginning of each day satisfies this condition, the device can power itself without any manual charging requirements. 
Hence, we introduce the following condition for finite horizon boundaries: 
\begin{equation} \label{eqn:targetEnergy}
E^B_{0} \geq E_{target} \hspace{2mm} 
\end{equation}

Figure~\ref{fig:block} summarizes the battery energy dynamics and the constraints presented above.
We also list the parameters and notation used throughout the work in Table~\ref{tab:notation}.

\vspace{1mm}
\noindent \textbf{Utility function:} 
We define the utility as the QoS that the target application provides to the user.
Although the proposed framework does not depend on any particular application, we demonstrate it on health monitoring and activity tracking applications. 
Any application has an energy requirement to run correctly, such as guaranteeing a certain number of measurements per unit time. 
We denote the minimum energy required by the target application by $M_E$. 
If we fail to allocate $M_E$ to the device, the device cannot provide useful output even if it is active.
In general, the device can deliver higher utility (i.e., better QoS) with larger allocated energy. 
However, allocating more energy has a diminishing rate of return 
since excessive energy does not necessarily improve the QoS. 
To capture this behavior, one can employ the following utility function:
\begin{equation}\label{eqn:log_utility}
u(E^A_t) = \alpha\ln\bigg(\frac{E^A_t}{M_E} \bigg)^\gamma
\end{equation}
where the parameters $\alpha$ and $\gamma$ are used to tune the utility function for a specific user or application.
Since this function is concave, it simplifies finding the optimal solution using the Karush-Kuhn-Tucker conditions~\cite{bhat2017near}.
However, such a generic utility function might not capture the behavior of 
the specific applications.
For example, Figure~\ref{fig:utility_illustration} shows the utility of different applications according to the energy consumption along with the generalized utility function in Equation~\ref{eqn:log_utility}.
Actual energy consumption and utility (measured by recognition accuracy) of real studies exhibit varying patterns~\cite{bhat2019reap,park2020energy}.
For example, the dashed lines in Figure~\ref{fig:utility_illustration} show data from two real applications, which significantly deviate from the logarithmic utility function.
To address this issue, the proposed approach works with both logarithmic and arbitrary utility functions, as described in Section~\ref{sec:nearoptgeneral}.

\begin{figure}[!t]
	\centering
	\includegraphics[width=0.7\linewidth]{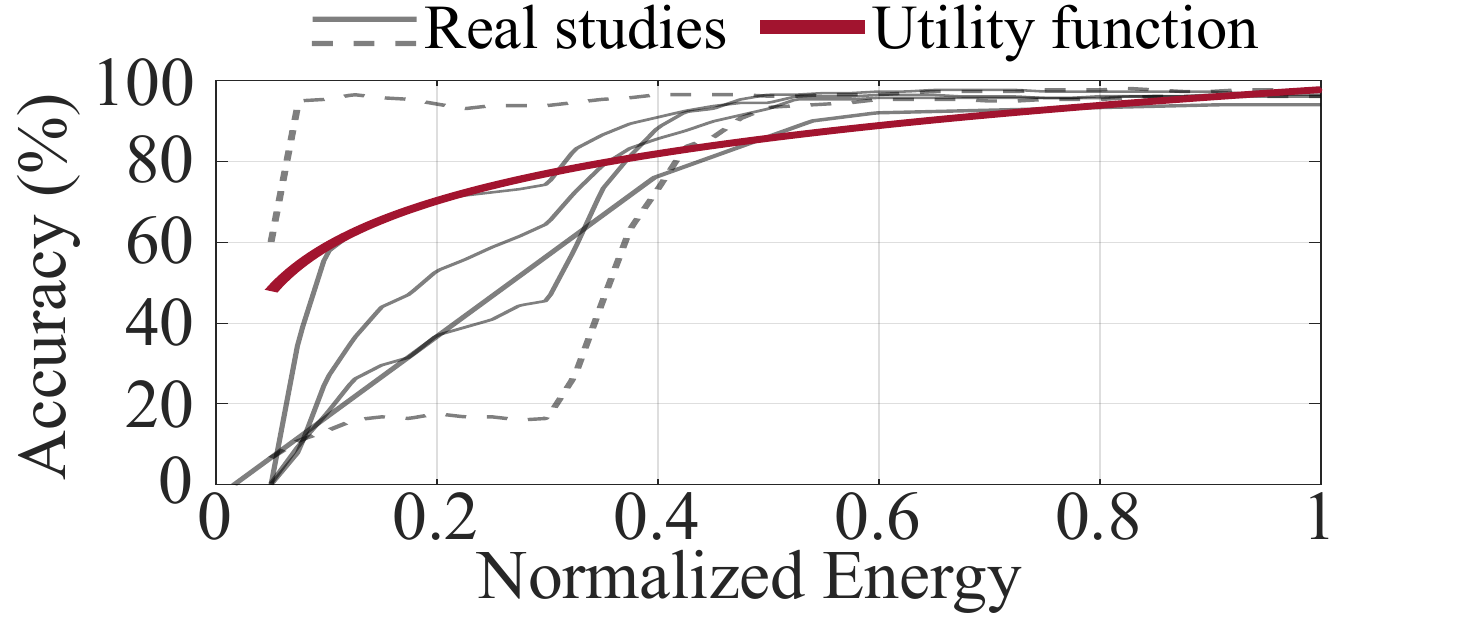}
	\caption{The generalized logarithmic utility function and utility curves from real studies. }
	\label{fig:utility_illustration}
	\vspace{-2mm}
\end{figure}

\section{Optimal Energy Management}
\label{sec:optem}

\subsection{Problem Formulation} \label{sec:formulation}

Our goal is to maximize the total utility over a finite horizon with the battery dynamics explained in Section~\ref{sec:overview} to guarantee an energy-neutral operation. 
The energy allocated to each interval can be expressed as a vector as: $\mathbf{\overline{E}^A} = \ECvec  \in \mathbb{R}^{T}$.
Similarly, let
$\mathbf{\overline{E}^H} = \hvec  \in \mathbb{R}^{T}$
be the extra energy added in each interval. 
Note that the first interval carries over the initial battery energy, 
while the rest contain only the energy harvested during that interval.
Finally, we express the lower-bound and upper-bound energy level 
to obtain a concise problem formulation:
$\mathbf{\overline{B}_L} = \bvec$, $\mathbf{\overline{M}_L} = \mvec \in \mathbb{R}^{T}$.
Using this notation, the optimization problem can be formulated as follows:
\begin{equation}
\label{eqn:opt_problem}
\begin{aligned}
& \text{$maximize$} & & \sum_{t=0}^{T-1}
\beta^t u(E^A_t) \\
& \text{$subject~to$} & & L(\mathbf{\overline{E}^{H}}-\mathbf{\overline{E}^A}) - \mathbf{\overline{B}_L} \geq 0 \\
&                     & & \rev{L(\mathbf{\overline{E}^{H}}-\mathbf{\overline{E}^A}) - \mathbf{\overline{M}_L} \leq 0}
\end{aligned}
\end{equation}
The optimization objective is the total utility expressed as a function of the energy allocations over the horizon 
and the discount factor $0 < \beta \leq 1$ that enables de-emphasizing future intervals. 
The set of constraints given in Equation~\ref{eqn:battery_energy}--Equation~\ref{eqn:targetEnergy} 
are written in matrix form 
using vectors $\mathbf{\overline{E}^{H}}$, $\mathbf{\overline{E}^A}$,  $\mathbf{\overline{B}_L}$, $\mathbf{\overline{M}_L}$, and a coefficient matrix 
$L \in \mathbb{R}^{T \times T}$:
\begin{equation}
\label{eqn:Lmatrix}
L = \Lmat \in \mathbb{R}^{T \times T}   
\end{equation}

\renewcommand{\arraystretch}{1}
\begin{table}[b!]
\caption{Notation table.}
\vspace{-2mm}
\label{tab:notation}
\begin{tabular}{@{}ll@{}}
\toprule
\textbf{Symbol}                                   & \textbf{Description}                                    \\ \midrule
$T$                                               & The finite time horizon                                 \\
$t$                                               & The time interval $[t,t+1)$                             \\
$\mathcal{S_T}$                                   & Set of uniform time intervals $t$ in the horizon $T$    \\
$E^A_t$, $E^H_t$                                  & Allocated and expected harvested energy in interval $t$          \\
$E^B_t$                                           & Battery energy at the beginning of interval $t$         \\
$E_{min}$, $E_{max}$                             & Minimum and maximum battery energy constraints           \\
$E_{target}$                                & Target battery energy constraint \\
$\widehat E^A_t$                                      & The corrected allocation at interval $t$          \\
$\widehat E^H_t$                                      & The actual harvested energy at interval $t$        \\
\rev{$\eta$}                                          & \rev{Parameter for harvesting and charging inefficiencies} \\
$\Delta^A_t$, $\Delta^H_t$ & \begin{tabular}[c]{@{}l@{}}Deviation from the expected values of allocated\\ 
                                                        and harvested energy in interval $t$\end{tabular}   \\
$\Delta_t$                                        & Cumulative deviation from the actual values until $t$   \\
$u(.)$, $\vec u(.)$                               & Utility function and vector valued utility function     \\
$M_E$                                             & Minimum energy required for positive utility            \\
$\alpha$, $\gamma$                                &  Parameters to tweak the shape of the utility function  \\ 
$\beta$                                           & Discounting factor for utility                          \\
$L$                                               & Coefficient matrix. Please see Equation~\ref{eqn:Lmatrix}  \\
$\mathbf{\overline{E}^A}$                         & $\ECvec  \in \mathbb{R}^{T}$                     \\
$\mathbf{\overline{E}^{H}}$                       & $\hvec   \in \mathbb{R}^{T}$                     \\
$\mathbf{\overline{E}^B}$                         & $\EBvec  \in \mathbb{R}^{T}$                     \\
$\mathbf{\overline{B}_L}$                         & $\bvec   \in \mathbb{R}^{T}$                     \\
\rev{$\mathbf{\overline{M}_L}$}                         & \rev{$\mvec   \in \mathbb{R}^{T}$}                     \\
\rev{$\lambda_1$, $\lambda_2$}                                         & \rev{Vectors of Lagrange multipliers}                          \\
$\nabla \vec u(.)$, $\nabla \vec   u^{-1}(.)$ & \begin{tabular}[c]{@{}l@{}}Gradient of the vector valued utility function \\ 
                                                                            and its inverse\end{tabular}    \\
$N$                                               & Number of nodes in rollout               \\
\rev{$\epsilon$}                                            & \rev{\begin{tabular}[c]{@{}l@{}}Parameter to set the interval of coefficients\\ 
                                                        in the rollout phase\end{tabular}  }\\
$\tilde U(.)$                                     & Heuristic function to compute future utility            \\
\bottomrule
\end{tabular}
\end{table}

\subsection{Optimal Solution using Iterative Gradient Projection} \label{sec:iterative}

The optimal solution to the problem given in Equation~\ref{eqn:opt_problem} 
can be obtained by writing the Lagrangian as: 
%
%
\begin{equation}
\label{eqn:lagrangian}
\begin{split}
    \rev{\mathcal{L} = \vec u(\mathbf{\overline{E}^A}) + \lambda_1^T(L\mathbf{\overline{E}^{H}} - L\mathbf{\overline{E}^A} - \mathbf{\overline{B}_L}) - } \\
    \rev{\lambda_2^T(L\mathbf{\overline{E}^{H}} -  L\mathbf{\overline{E}^A} - \mathbf{\overline{M}_L})}
\end{split}
\end{equation}
\rev{where $\lambda_1, \lambda_2 \in \mathbb{R}^T$ are the Lagrange multipliers} and $\vec u(\mathbf{\overline{E}^A}) \colon \mathbb{R}^T \to \mathbb{R}^T$ 
is a vector function defined as follows: $\uvec$. 
Using the Lagrangian, we can write the following Karush-Kuhn-Tucker conditions~\cite{kuhn1951}:
\begin{align}
\label{eqn:conditions}
    & \rev{\nabla \vec u(\mathbf{\overline{E}^A}) - (\lambda_1^T-\lambda_2^T)L = 0 }\\ \nonumber
    & \rev{ \lambda_1^T(L\mathbf{\overline{E}^{H}} - L\mathbf{\overline{E}^A} - \mathbf{\overline{B}_L}) = 0, \quad \lambda_1 \geq \mathbf{\overline{0}} }\\
    & \rev{ \lambda_2^T(L\mathbf{\overline{E}^A} - L\mathbf{\overline{E}^{H}} + \mathbf{\overline{M}_L}) = 0, \quad \lambda_2 \geq \mathbf{\overline{0}} } \nonumber
\end{align}
where $\mathbf{\overline{0}} \in \mathbb{R}^T$ is a vector of zeros and $\nabla \vec u(\mathbf{\overline{E}^A}) \colon \mathbb{R}^T \to \mathbb{R}^T$ is the gradient of $\vec u(\mathbf{\overline{E}^A})$ defined as follows: $\graduvec$.

\vspace{2mm}
\noindent \textbf{Challenge:}
Finding the optimal solution is challenging in real application scenarios for two reasons. 
First, the set of equations in Equation~\ref{eqn:conditions} does not have a closed-form solution for an arbitrary utility function. 
Second, the solution requires the energy values that will be harvested in 
\textit{future time intervals}.

\vspace{1mm}
\noindent \textbf{Finding the optimal solution offline:}
Due to the challenges mentioned above, 
the optimal energy allocation can only be obtained 
using iterative algorithms and \textit{expected values} of the 
energy that will be harvested in future intervals. 
\textit{To obtain the optimal solution as point of reference}, 
we adapt an iterative gradient projection algorithm 
used recently in network utility maximization problems~\cite{karakocc2020multi}.
\rev{It starts with random allocations and Lagrange multipliers $\lambda_1, \lambda_2$, 
as depicted in Algorithm~\ref{alg:iter} lines 3--5.  
At each iteration, the algorithm finds the optimal allocations $\mathbf{\overline{E}^A}$ 
using the inverse of the gradient function  
and the latest Lagrange multipliers (line 8). 
Then, the Lagrange multipliers are updated 
following a feasible descent direction 
given by the optimization constraints (lines 9--10).
The iteration in lines 6--11 continues until $\lambda_1, \lambda_2$ converge
within desired tolerance bounds. }
The proof of convergence can be found in~\cite{karakocc2020multi}. 

\vspace{1mm}
\noindent \textbf{The need for a runtime solution:}
The actual energy harvested in each time interval differs from the expected values used by offline algorithms.  
Therefore, the energy allocations must be recomputed  after each interval to account for these deviations. 
Iterative algorithms cannot be used repeatedly for this purpose due to their significant runtime overhead. 
For example, the execution time of iterative gradient projection algorithm (Algorithm~\ref{alg:iter}) is 70.3 seconds on TI CC2652R microcontroller~\cite{ticc2652}, as detailed in Section~\ref{sec:expresults}. 
In contrast, the light-weight energy allocation algorithm presented in~\cite{bhat2017near} takes only 2~ms, resulting in over four orders of magnitude energy consumption savings, albeit with a significant loss in optimality. 
Since significant energy overhead is prohibitive, 
there is a \textit{strong need for runtime solutions that can approach the performance of the iterative algorithms.}
The following sections present our runtime solutions to the optimization problem in Equation~\ref{eqn:opt_problem}. We start with a constant time solution for logarithmic utility function. Then, we present a novel rollout algorithm for arbitrary utility functions.

\vspace{-2mm}
\subsection{Light-Weight Algorithm for Logarithmic Utility Functions}
\label{sec:runtime}

The energy allocation problem given in Equation~\ref{eqn:opt_problem} does not have a closed-form solution for two reasons: 
1) Arbitrary shape of the utility function,
\rev{2) The minimum and maximum energy constraints in Equation~\ref{eqn:minEnergy}.
One can derive a closed-form expression by using a logarithmic utility function and 
relaxing the minimum and maximum energy constraints.}
When the utility function is defined as 
$u(E^A_t) = \ln(\frac{E^A_t}{M_E})$, 
the optimal energy allocations are given by:
\begin{align} 
\label{eqn:EC0}
\begin{tabular}{rp{1mm}}
    \begin{tabular}[c]{@{}r@{}}First\\ interval\end{tabular} & \begin{tabular}{l} \hspace{-2mm}:\end{tabular}
\end{tabular}
&E^A_0 =\frac{E^B_0 - E_{target} + \sum_{t=0}^{T-1}E_t^H} {1+\beta+\beta^2+\ldots+\beta^{T-1} } \\
\begin{tabular}{rp{1mm}}
    \begin{tabular}[c]{@{}r@{}}Subsequent\\ intervals\end{tabular} & \begin{tabular}{l} \hspace{-2mm}:\end{tabular}
\end{tabular}
&E^A_{t+1} =\beta E^A_t  ~\hspace{5mm}  0 \leq t \leq T-1 \nonumber
\end{align}
where $E_t^H$ is the energy expected to be harvested in interval $t$ and $\beta$ is the discount factor~\cite{bhat2017near}. 

\begin{figure}[t]
    \setcounter{figure}{0}
    \makeatletter 
    \renewcommand{\figurename}{Algorithm}
    \makeatother
    \centering
    \caption{Iterative gradient projection algorithm.}
    \label{alg:iter}
    \vspace{1mm}
    \begin{algorithmic}[1]
    \small
        \STATE \textbf{Input:} \rev{ $\mathbf{\overline{E}^A}$ and $\lambda_1^*, \lambda_2^*$ }
        \STATE \textbf{Output:} \rev{ $\mathbf{\overline{E}^A}$ and $\lambda_1^*, \lambda_2^*$ }
        \STATE $\mathbf{\overline{E}^A} \gets rand(T,1)$\
        \STATE \rev{$\lambda_1, \lambda_2 \gets rand(T,1)$}\
        \STATE \rev{$\lambda_1^*, \lambda_2^* \gets rand(T,1)$}\
         \WHILE{\rev{$||\lambda_1^* - \lambda_1|| >$ tolerance \textbf{and} $||\lambda_2^* - \lambda_2|| >$ tolerance} }
              \STATE \rev{ $\lambda_1 \gets \lambda_1^*$\, $\lambda_2 \gets \lambda_2^*$}\
              \STATE \rev{$\mathbf{\overline{E}^A} \gets \nabla \vec u^{-1}(\lambda_1^T-\lambda_2^T)L$}\ (follows  from Equation~\ref{eqn:conditions})
              \STATE \rev{$\lambda_1^* \gets$ max\{$\lambda_1 + stepSize\times(L\mathbf{\overline{E}^A}-L\mathbf{\overline{E}^{H}}+\mathbf{\overline{B}_L}), \overline{0}$\} }
              \STATE \rev{ $\lambda_2^* \gets$ max\{$\lambda_2 + stepSize\times(L\mathbf{\overline{E}^{H}} - L\mathbf{\overline{E}^A} - \mathbf{\overline{M}_L}, \overline{0}$\} }
              ($stepSize=0.001$ in our implementation)
          \ENDWHILE
    \end{algorithmic}
    \vspace{-4mm}
    \normalsize 
    \setcounter{figure}{2}
\end{figure}

\vspace{1mm}
\noindent \textbf{Correcting the initial allocations:} 
The initial energy allocations are found at the beginning of the finite horizon using Equation~\ref{eqn:EC0}. 
These allocations must be corrected to account for the difference between the actual and expected energy values. 
Let $\Delta^H_t$ denote the difference between the expected and actual energy harvested 
in interval $t \in \mathcal{S_T}$. 
Similarly, let $\Delta^A_t$ be the difference between the initial and corrected allocated energy in the same interval. 
Using these definitions, the cumulative deviation from the actual values can be expressed as: 
\begin{equation} \label{eqn_Deltat}
\Delta_t = \sum_{\tau=0}^{t-1} (\Delta^H_\tau + \Delta^A_\tau)    
\end{equation}
This differences must be compensated at runtime by perturbing the initial allocations since they can lead to under- or over-utilization of harvested energy. 
The following lemma gives the re-allocations that maintain the optimality.
\begin{lem}
\label{lem:correction}
The initial allocation $E^A_t~t \hspace{-1mm}\in \hspace{-1mm} \mathcal{S_T}$ must be perturbed according to the following equation to redistribute the energy 
 surplus/deficit and maintain the optimality.
\begin{equation} \label{eqn_Ecorrection}
    \widehat E^A_t = E^A_t  + \Delta_t \frac{(1-\beta)}{1-\beta^{T-t}}
\end{equation}
\end{lem}
\noindent
In Equation~\ref{eqn_Ecorrection}, $\widehat E^A_t$ denotes the corrected allocation at interval $t$.
Note that $\Delta_t$ defined in Equation~\ref{eqn_Deltat} represents 
the cumulative deviation until the current interval, 
where positive and negative values imply a surplus or deficit, respectively.
Hence, a portion of the cumulative deviation is added to the initial allocation.
The derivation of Lemma~\ref{lem:correction} is given in Appendix~\ref{sec:app1}.

\rev{Equation~\ref{eqn_Ecorrection} can cause the battery to drain below $E_{min}$ or rise above $E_{max}$ since these constraints were relaxed to find a closed-form solution. 
Hence, the final consideration is enforcing the minimum and maximum energy constraints at runtime.
To achieve this, we project the remaining battery energy $E^B_{t+1}$ at runtime using Equation~\ref{eqn:battery_energy}, and compare it against $E_{min}$ and $E_{max}$ before committing to a solution. 
If there is a minimum energy constraint violation, we allocate the maximum energy that satisfies $E^B_{t+1}$ = $E_{min}$. 
Similarly, if there is a maximum energy constraint violation, we allocate the excess energy such that $E^B_{t+1}$ = $E_{max}$.}
That is, the allocation becomes:
\begin{equation} 
\label{eqn:Ec_update}
\widehat E^A_t = \begin{cases}
\rev{ E^A_t + \Delta_t \frac{1-\beta}{1-\beta^{T-t}} } & \rev{E_{max} \geq E^B_{t+1} \geq E_{min} } \vspace{2mm} \\
\rev{ E^A_t + (E^B_{t+1} - E_{max}) } & \rev{ E^B_{t+1} > E_{max} }\vspace{2mm} \\ 
\rev{ E^B_t + E^H_t - E_{min}  } & \rev{ E^B_{t+1} < E_{min} }
\end{cases}
\end{equation}

\noindent \textbf{Insight:}
When the battery energy is not scarce (i.e. $E_{max} \geq E^B_t>>E_{min}$), 
the solution obtained by Equation~\ref{eqn:Ec_update} is optimal.
Otherwise, the piecewise nonlinearity in Equation~\ref{eqn:Ec_update} is triggered and the solution becomes sub-optimal.

\vspace{-2mm}
\subsection{Runtime Rollout Algorithm for Arbitrary Utility Functions}
\label{sec:nearoptgeneral}

The energy allocation technique presented in the previous section is valid 
only for the logarithmic utility function defined in Section~\ref{sec:overview}.
In real applications, the ``utility'' is often represented as measured metrics such as accuracy or throughput, which may not follow a logarithmic trend with the allocated energy, as shown in Figure~\ref{fig:utility_illustration}.
In fact, target devices may have a finite set of operating points
with different energy consumption vs. utility trade-offs~\cite{bhat2019reap}. 
In these cases, the allocation algorithm should choose the operating points that maximize 
the accuracy under the overall energy harvesting budget over the horizon.
Therefore, practical energy allocation algorithms must support arbitrary and discrete utility functions.

\begin{figure}[b]
\vspace{-4mm}
\centering
\includegraphics[width=0.73\columnwidth]{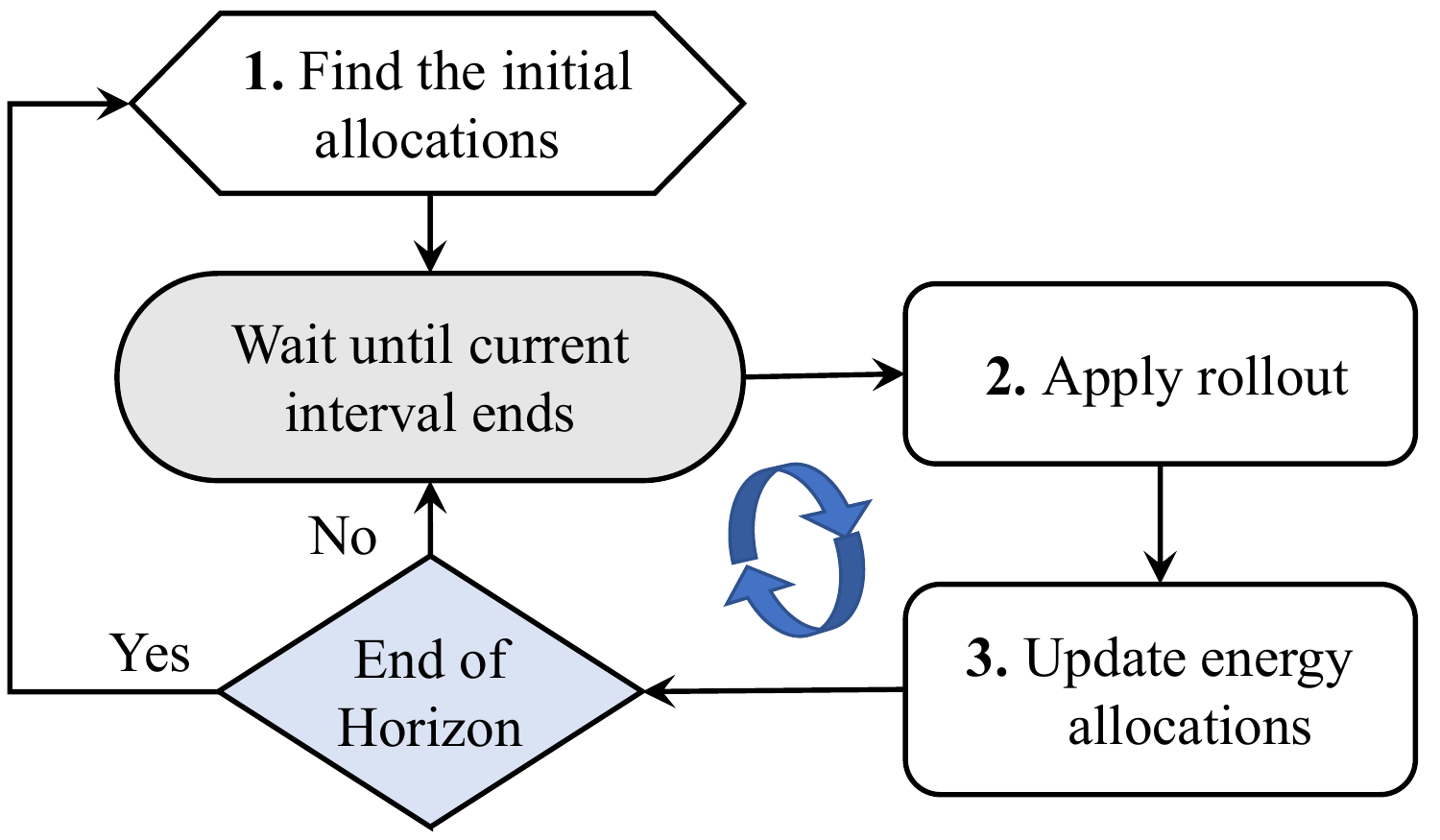}%
\caption{The overview of the proposed rollout algorithm.}
\label{fig:alg_overview}
\end{figure}

This section presents a rollout algorithm to achieve optimal runtime energy allocation under 
arbitrary utility functions. 
To achieve this objective, the proposed algorithm first finds the initial allocations 
at the beginning of the finite horizon, as illustrated in Figure~\ref{fig:alg_overview}. 
At the end of each interval, it determines the deviations from the actual and expected energy values (i.e., $\Delta_t^H$ and $\Delta_t^A$). 
Then, it uses this information to perform a rollout phase before the start of the next time interval.
Finally, it updates the energy allocations and starts the next interval. 
These three steps are repeated until the end of the finite horizon, as shown in Figure~\ref{fig:alg_overview}. 
The rest of this section describes each of these steps in detail.

\subsubsection{Finding the Initial Allocations}
The first step of ECO is computing 
the initial energy allocations in the beginning of the time horizon. 
This can be done using any technique, including the closed-form solution given in Equation~\ref{eqn:EC0}. 
In this work, we employ the iterative gradient projection algorithm 
presented in Section~\ref{sec:iterative} for two reasons. 
First, it supports arbitrary utility functions. 
Second, its runtime overhead can be hidden since it runs only once per finite horizon (e.g., a day) 
using the expected energy values. 
Since the subsequent rollout phase corrects the initial allocations at runtime, we can increase the tolerance threshold in Algorithm~\ref{alg:iter} without decreasing the utility considerably.
For example, the implementation on our hardware prototype takes 0.993 seconds to find the initial allocations. 
This execution time amounts to only 0.03\% of a one-hour interval (or 1.6\% of a one-minute interval). 
Combined with the efficient rollout phase that runs every interval, the total overhead is less than 0.1\% of running Algorithm~\ref{alg:iter} in each interval.
At the end of this step, we obtain the initial allocations $\mathbf{\overline{E}^A}$.

\subsubsection{The Rollout Phase}

\begin{figure}[!h]
\centering
\vspace{-2mm}
\includegraphics[width=0.45\textwidth]{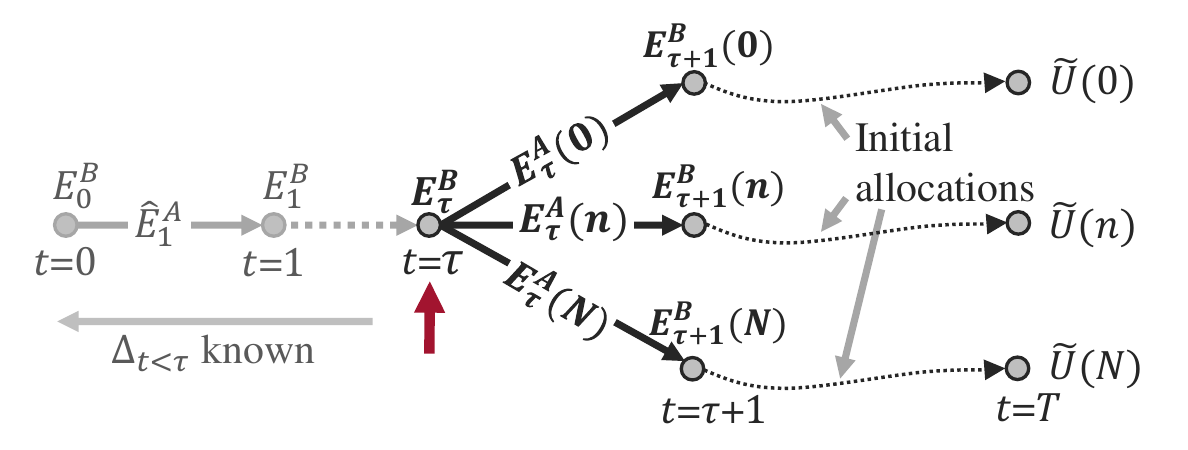}%
\vspace{-3mm}
\caption{Illustration of the rollout phase.}
\label{fig:rollout}
\vspace{-2mm}
\end{figure}

\rev{At the end of each interval, ECO adjusts the energy allocations using the cumulative difference between the expected and actual energy values, i.e., $\Delta_t$ defined in Equation~\ref{eqn_Deltat}. Suppose that the current time interval is $t = \tau$, as illustrated in Figure~\ref{fig:rollout}. 
ECO starts with the initially allocated energy and explores different re-allocation strategies represented by the branches. 
The first branch results in the smallest allocation, while the consecutive branches allocate a linearly increasing amount of energy.} 

\rev{To quantify the linearly increasing energy allocation illustrated by the branches in Figure~\ref{fig:rollout}, our rollout strategy replaces the $\frac{1-\beta}{1-\beta^{T-t}}$ term in Equation~\ref{eqn:Ec_update} with a coefficient as follows:}
%
%
\begin{equation} 
\label{eqn:Ecbranch}
E^A_\tau(n) = \mathbf{\overline{E}^A}[\tau] + \Delta_\tau (1 - \epsilon + \frac{2\epsilon}{N}n),~0 \leq n < N
\end{equation}
Each different value of $n$ leads to a different reallocation, i.e., branch, as illustrated in Figure~\ref{fig:rollout}.
The parameter $N$ can be the number of operating points or can be a finite number (e.g., 10) that controls the exploration space.
Then, the algorithm follows each branch and uses the initial allocations as the base policy for the rest of the intervals to find the expected utility. 
The total future utility at the end of the horizon is computed as follows:
\begin{equation} 
\label{eqn:heuristic}
\tilde U(n) = \begin{cases}
-\infty   & E^B_{t>\tau}(n) < E_{min} \\
\rev{ -\infty  } & \rev{ E^B_{t>\tau}(n) > E_{max} } \\
-\infty   & E^B_{T}(n) \leq E_{target} \\
\sum_{t=\tau+1}^T u(\mathbf{\overline{E}^A}[t]) & otherwise 
\end{cases}
\end{equation}
%
The first two conditions check if the allocation violates the minimum and maximum battery level constraints during the remaining time intervals. 
The third condition checks the target battery energy constraint at the end of the day.
These checks eliminate aggressive allocations and hence reduce over-allocation.

\begin{figure*}[!b]
\vspace{-6mm}
\centering
\subfloat[]{\includegraphics[width=0.49\textwidth]{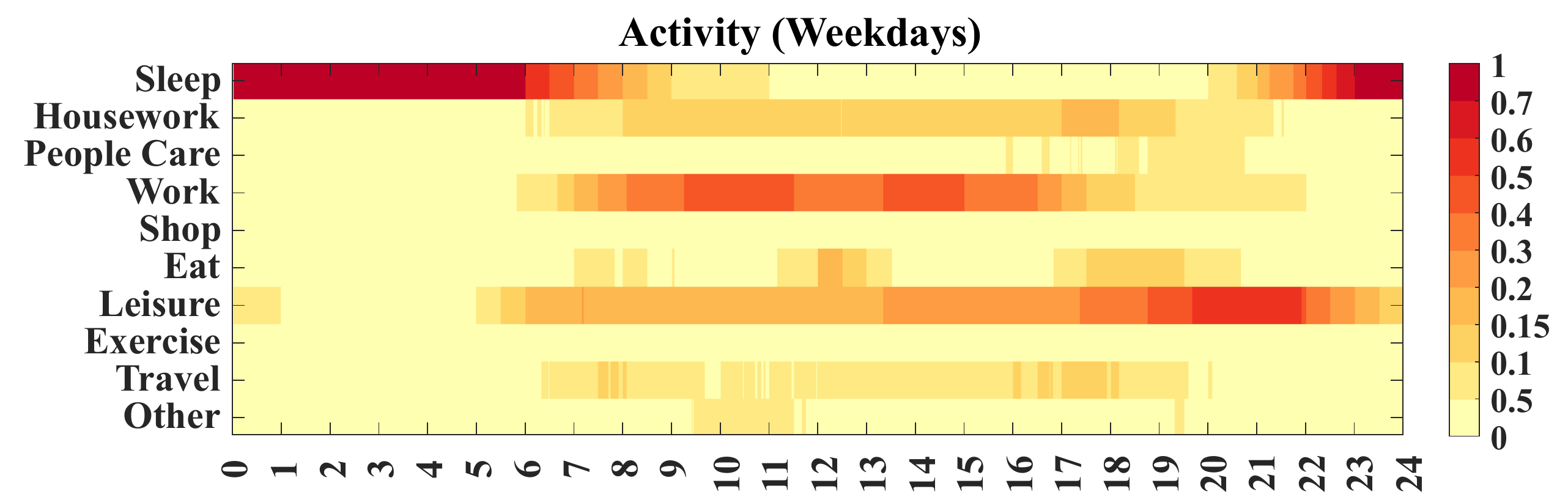}%
\label{fig:ATUS_1}
}
\hfil
\subfloat[]{\includegraphics[width=0.49\textwidth]{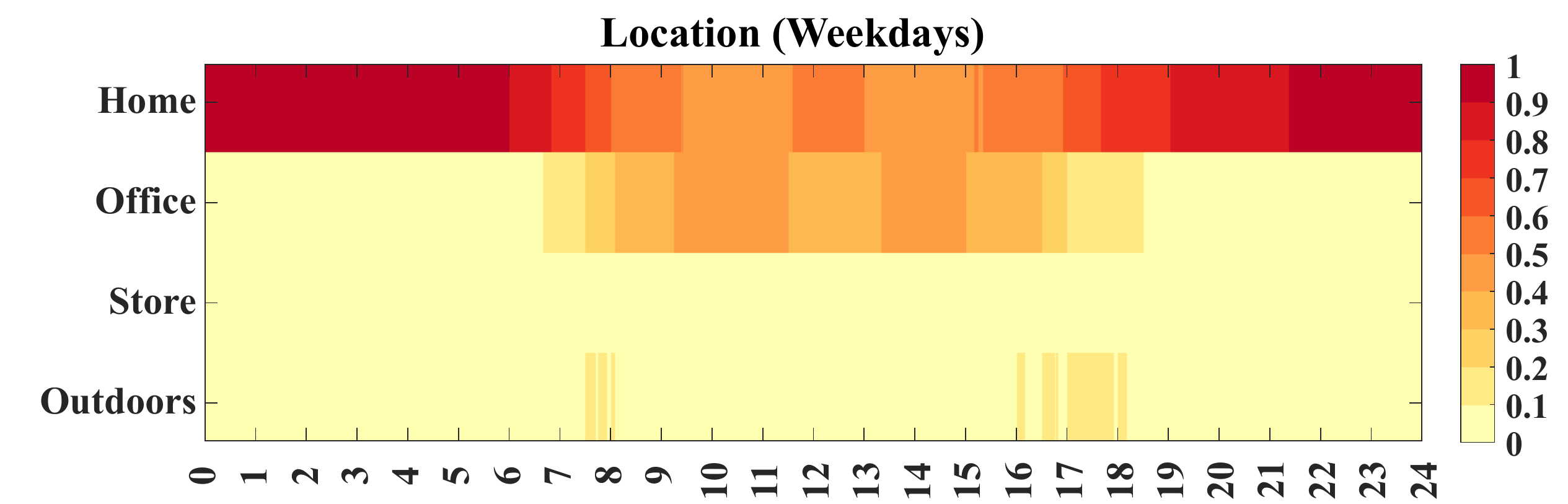}%
\label{fig:ATUS_2}
}
\caption{Percentage of 4772 users a) doing different activities b) at different locations throughout the day.}
\label{fig:ATUS}
\end{figure*}

\subsubsection{Updating the Energy Allocations}

The rollout phase calculates the expected utility achieved by each branch.
Using this information, the revised energy allocations $\widehat E^A_t$ are obtained from a modified version of Equation~\ref{eqn:Ec_update} as follows:
\begin{equation} 
\label{eqn:Ec_rollout}
\widehat E^A_t = \begin{cases}
\begin{aligned}
& \underset{E^A_t(n)}{\text{argmax}}~u\big(E^A_t(n)\big) + \tilde U\big(n\big),  && E^B_{t+1} \geq E_{min} \vspace{2mm} \\ 
 &   \rev{\underset{E^A_t(n)}{\text{argmax}}~u\big(E^A_t(n)\big) + \tilde U\big(n\big) +}&& \hspace{-3mm}  \rev{(E^B_{t+1}  - E_{max}),} \\
    & && \rev{E^B_{t+1} > E_{max} }
 \vspace{2mm}  \\
&E^B_t +  E^H_t - E_{min},  && otherwise 
\end{aligned}
\end{cases}
\end{equation}
where $u(.)$ is the utility function and $\tilde U(.)$ is the heuristic that gives the future utility when the base policy is applied.
If the battery constraints are not violated, the algorithm follows the arc ($E^A_t(n)$) that maximizes the sum of immediate utility and future utility $u\big(E^A_t(n)\big) + \tilde U\big(n\big)$.
\rev{If the device violates the minimum energy constraint during this period, it will go into a low-power state to conserve energy.}
Otherwise, the battery constraints are enforced as explained in Equation~\ref{eqn:Ec_update}.
\rev{This is done by first doing a linear descending sort of the values of sum and then checking the constraints.}
This sequential optimization technique guarantees an improvement over the base policy it is initially supplied with~\cite{bertsekas2010rollout}.

\section{Energy Harvesting Model}
\label{sec:EHModel}

This work employs a combination of light and motion energy as the ambient energy sources and a flexible Li-ion battery as the energy storage unit. 
We use the ATUS dataset~\cite{amtus} to extract the location and activity patterns of 4772 users. 
Then, we apply our light and motion EH models to the extracted data and generate 24-hour EH estimates per user. 
In what follows, we describe the data pre-processing steps we apply to the ATUS dataset and the EH models.
Finally, we present example EH values from different users to illustrate the final EH model.

\vspace{-2mm}
\subsection{ATUS 2018 dataset pre-processing}
ATUS conducted by the US Department of Labor~\cite{amtus} is a comprehensive dataset that contains the amount of time people spend doing various activities. 
The dataset consists of daily activity patterns of 9593 users with 18 different categories of activities, such as personal care, working, or eating. 
4772 users completed the survey on weekdays, whereas the remaining 4821 users completed it during weekends. 
We reduce the number of activity labels to 10 by grouping the infrequent activities, such as religious activities and eldercare, under the title ``Other".
The precise mapping of ATUS activity labels to our labels is provided in Appendix~\ref{sec:app2}.
Figure~\ref{fig:ATUS_1} shows the reduced set of activity labels and the percentage of 4772 users doing a particular activity throughout the day. 

The location information is required to estimate the amount of energy harvesting from the light because the irradiance affects light EH efficiency.
We assign a location label from the set of \{`home',`office',`store',`outdoors'\} to each activity following common sense.
For example, sleep activity is assigned `home' as the location, `work' is assigned `office', `shopping' is `store', `eat' is assigned `home' or `office' depending on time of day. 
The mapping of location labels to activities is provided in Appendix~\ref{sec:app2}.
Figure~\ref{fig:ATUS_2} shows the percentage of 4772 users being at the corresponding locations as a function of time. 
As a result, we obtain hourly activity and location information per user for a typical day in their lives.

\begin{figure*}[!t]
\centering
\includegraphics[width=0.9\textwidth]{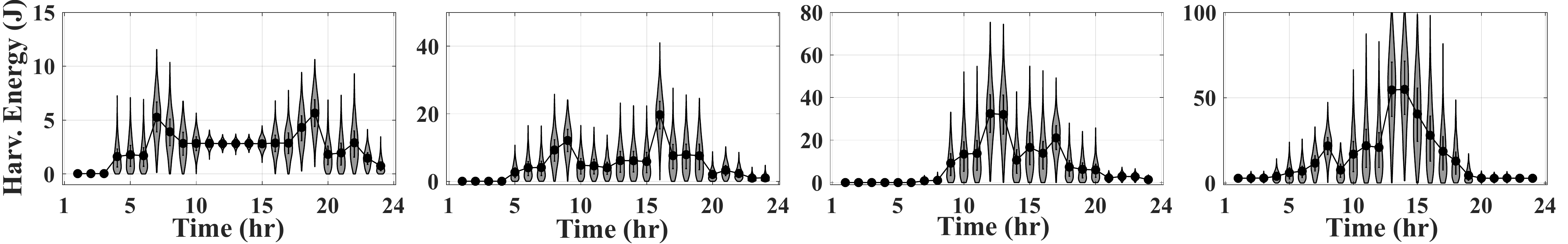} 
\vspace*{-4mm}
\caption{\rev{EH models of the median users from the four clusters.}}
\vspace*{-5mm}
\label{fig:EHMODEL}
\normalsize
\end{figure*}

\vspace{-2mm}
\subsection{Light energy harvesting model}
A light energy harvester uses a photovoltaic cell (PV-cell) to transduce the environment's light energy into usable electrical power. 
FlexSolarCells SP3-37, a lightweight, flexible, and small form-factor PV-cell, is used as the transducer~\cite{sp3-12}. 
We identified the I-V characteristics of SP3-37 in a previous study through a set of controlled experiments with a halogen lamp~\cite{park2017flexible,tuncel2020special}. 
This information is used to compute the harvested light energy when the irradiance is known.

PV-cells generate relatively high power in outdoor lighting conditions (over 1~mW/cm$^2$) where the irradiance is high~\cite{chong2019energy}. 
To obtain the outdoors irradiance, we first estimate the sun's position at a given date and time using Sandia’s Ephemeris model~\cite{Sandia}. 
Then, we convert the position information to radiation using Ineichen’s model~\cite{ineichen2002new}. 
The output corresponds to the irradiance around the user when the user is outdoors.

The energy generated by a PV-cell quickly reduces 
for indoor lighting conditions where the irradiance is lower~\cite{tuncel2020special}. 
Furthermore, previous studies showed that indoor lighting conditions at the home, office, and store environments significantly differ from each other. 
The work in~\cite{wang2008design} provides measurements of irradiance in a typical office lighting condition. 
Besides, the information from~\cite{Flex,lightlevel} shows that on average, store lighting intensity is 3$\times$ of office lighting and home lighting is 0.5$\times$ of office lighting. 
We use this information to find the irradiance when the user is at home, at an office, or at a store. 
Therefore, we combine the hourly location information obtained from the ATUS dataset, the irradiance amounts, and the I-V characteristics of SP3-37 to estimate hourly light EH per user.

\vspace{-2mm}
\subsection{Motion energy harvesting model}
A piezoelectric energy harvester transduces mechanical energy into usable electrical power. 
We use joint bending as the source of mechanical energy, which is enabled by advances in flexible piezoelectric transducers, such as Polyvinylidene Fluoride (PVDF) and Macro-Fiber Composite (MFC). 
A recent study analyzes piezoelectric EH from joint bending motion and shows that 7.8 $\mu$J per step can be harvested from the knee with an MFC8528P2 piezoelectric element during walking~\cite{tuncel2020towards}. 
We use this knowledge to compute the harvested motion energy from the user's activity information. 

Piezoelectric transducers generate higher power with increasing piezoelectric material~\cite{tuncel2020towards}. 
Thus, we can use larger piezoelectric elements or connect multiple components in parallel to increase harvested power. 
Furthermore, MFC piezoelectric elements can be stacked on top of each other as long as they do not cause user discomfort. 
Following this, we consider ten elements per knee, where five MFC8528P2 elements lie across the knee, and another five sits on top. 
\rev{Low-effort walking with sporadic stepping is defined as 20-39 steps/min~\cite{tudor2018fast}. Using the middle value in this interval, we obtain 30 steps/min$\times$60 mins = 1800 steps/hour for activities that involve walking.}
We combine this with the activity information obtained from the ATUS dataset and compute 7.8 $\mu$J/step$\times$1800 steps/hour$\times$10 = 0.14 J for activities that involve walking. 
For the sake of simplicity, we assume 2$\times$ the walking energy for activities that involve exercise and no harvesting for activities with little to no movement, such as working and sleeping.

\vspace{-3.5mm}
\subsection{Combined light and motion energy harvesting model}
\vspace{-0.5mm}
We combine the hourly light and motion EH values by superimposing them. 
For example, consider a user who is performing an exercise outdoors between 3-4~PM. 
The light EH from the outdoor irradiance at 3~PM and motion EH from exercising are added together to find the expected EH between 3-4~PM.
This operation is performed for each hour to generate the 24-hour EH of a user. 
These values constitute the expected EH for a user, as shown with black markers on Figure~\ref{fig:EHMODEL}. 
\rev{The user may deviate from their expected day, causing differences between the expected and actual harvested energy. 
In addition, there could be other variations due to partial shading, dirt, or scratches on the PV-cells for light energy harvesting and variations due to inappropriate placement, misalignment, and sliding of the piezoelectric material on the knee for motion energy harvesting. 
Such occasions can have a negative or a positive impact on the amount of harvested energy.
For example, a misplaced PV-cell could receive more direct but less light due to partial shading by a piece of clothing. 
Thus, any other day of the same user is generated from the expected EH with random variations.
We add deviations from the expected EH values to model random variations by considering each hour as a rectified Gaussian distribution.
}
The variance of each distribution is found by a moving variance filter with a window size of 7. 
For example, the variance at 2~PM is computed by taking the moving variance of EH values between 11~AM to 5~PM. 
This approach inherently accounts for sleeping, as illustrated with the low-variance levels in the early hours of the day in Figure~\ref{fig:EHMODEL}.
We also introduce a normalization factor to the model to control the maximum standard deviation in the day.
In Section~\ref{sec:expresults}, we sweep this factor from 0\% to 25\% to observe the effect of the uncertainty in EH on performance.
Finally, we do this procedure for every user in the dataset and obtain 4772 EH models in total. 

\rev{ 
Clustering the ATUS dataset into user groups can help analyze the user behavior and use the proposed ECO framework for new users. 
To this end, we first divide each day into three intervals: Midnight--8AM, 8AM--4PM, 4PM--midnight. 
Then, we added the energy harvested in each of the three intervals to reduce the data size to 4772$\times$3. 
Finally, we apply k-medoids clustering over these three features to obtain four (k=4) well-separated user groups. Figure~\ref{fig:EHMODEL} shows the EH models of the median users in each cluster. 
All four clusters have typically low (0-6 J) harvested energy in the first eight and last eight hours of the day. 
The first cluster has the least expected energy (less than 8 J) in the middle eight hours, whereas the fourth cluster harvests more than 25 J during the same period, and the other clusters fall in between.}

\rev{We envision that a new user will be placed in one of these clusters according to their energy harvesting usage pattern. 
When a new user starts using the device, we will log the harvested current going into the battery with the help of a coulomb counter and begin generating a coarse model. 
This coarse model will then be used to place the user in a cluster. 
Then as the user continues to use the device, their energy harvesting pattern is updated in time. 
We demonstrate the performance of ECO with varying degrees of model accuracy, from very accurate to very coarse, in Section VI.}

\section{Experimental and Simulation Results}
\label{sec:eval}
This section evaluates the performance of the proposed ECO framework and compares it to the prior work in the literature.
We demonstrate the proposed framework on a wearable hardware prototype designed for health and activity monitoring as the driver applications.
However, we stress that the proposed approach does not depend on any particular application and can be applied to any energy harvesting device.

\begin{figure}[!t]
	\centering
	\includegraphics[width=0.82\linewidth]{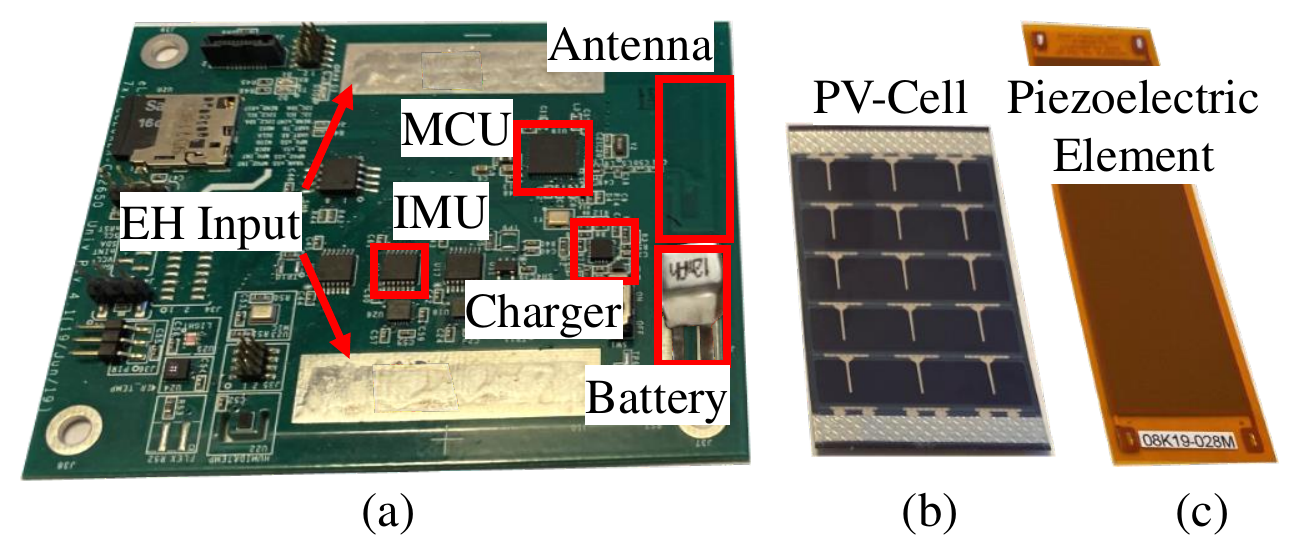}
	\vspace{-3mm}
	\caption{a) IoT Device Prototype b) SP3-37 c) MFC8528P2}
	\label{fig:prototype}
	\vspace{-3mm}
\end{figure}

\vspace{2mm}
\noindent \textbf{Experiment parameters:} 
We employ the prototype shown in Figure~\ref{fig:prototype} to demonstrate the proposed ECO framework under realistic scenarios. It consists of an MPPT charger (TI BQ25504~\cite{BQ25504}), a microcontroller (TI CC2652R~\cite{ticc2652}), an inertial measurement unit (ICM-20948~\cite{icm20948}), and various additional sensors that can be useful for wearable healthcare applications.
\rev{
The standby leakage currents of all components in our prototype device add up to 54.6 $\mu$A, which gives the total leakage energy in an hour as 0.64 J (3.3V$\times$54.6$\mu$A$\times$3600s).
Similarly, we keep the microcontroller ($\mu$C) in the sensor controller state, and activate it and the nonvolatile ram (NVram) only intermittently. 
Therefore, we calculate two different sums of active currents: The active currents of the sensors and power circuitry add up to 1.3 mA, while those of the $\mu$C and NVRam add up to 4.7 mA. 
Using these values, we obtain the total active energy of the sensors in an hour as 15.8 J with 100\% duty cycle, and total active energy of the $\mu$C and NVRam as 55.8 J with 100\% duty cycle.
However, the $\mu$C and NVRam do not need to be active all the time. 
We consider 30 seconds of active time in an hour sufficient for them, which makes their duty cycle 30/3600~=0.8\%. 
This scales the total active energy of the $\mu$C and NVRam down to 0.44 J in an hour. 
We obtain $M_E$ as 0.64 J + 0.44 J$=$1.08 J, which is the minimum amount of energy required for the device without any useful output. 
The rest of the harvested energy is used to duty cycle the sensors. 
For example, if 11.08 J of energy is harvested in an hour, 10 J of it can be used to turn on the sensing circuitry with a 10/15.8$=$63\% duty cycle, provided that battery constraints are not violated (and will not be in the future intervals). 
Finally, we can claim that if a total of 24$\times$1.08 J$=$25.9 J is harvested throughout the day, energy-neutrality can be achieved with our prototype device.}

\rev{Given the analysis above, Figure~\ref{fig:EHMODEL} shows that the amount of harvested energy is well above the required 25.9 J threshold for some users in the dataset. 
However, for some users it can be just above or even below the threshold. 
In the former case, the utility will be very near zero even though energy-neutrality is achieved. In the latter case, energy-neutrality can not be achieved. 
In fact, Figure~\ref{fig:res2}-2b shows that if the initial battery energy is scarce, close to 5\% of the users in the dataset cannot achieve energy-neutrality without violating the minimum energy constraint.
}

We use a 12 mAh Li-Po battery GMB 031009~\cite{031009} as the energy storage element. 
The IoT device parameters, such as $E_{min}$ and $M_E$, are obtained from this prototype and listed in Table~\ref{tab:exp_parameters}. 
The amount of energy stored in the battery is 160 J when it is full.
We consider 10\% and \rev{90\%} of the battery capacity as scarce energy level and abundant energy level, respectively.
\rev{The user’s expected location and activities and the environment around them may deviate from the expected day, causing significant differences between the expected and actual harvested energy. 
To observe the effect of such modeling errors, we sweep the standard deviation in the generated energy harvesting patterns from 0\% to 25\% with 5\% intervals. 
This amounts to a modeling mean absolute percentage error (MAPE) of 0\% to 60\% with 12\% intervals, respectively. 
Therefore, at higher standard deviation settings, the expected EH pattern is only coarsely accurate.}

\renewcommand{\arraystretch}{1.1}
\begin{table}[!t]
\centering
\caption{Parameter values used during evaluations}
\vspace{-2mm}
\label{tab:exp_parameters}
\begin{tabular}{lllll}
\toprule
\textbf{Parameter} & \textbf{Value} &  & \textbf{Parameter} & \textbf{Value} \\ \cline{1-2} \cline{4-5} 
$E_{min}$          & 10 J           &  & $M_E$              & 1.08 J         \\ \cline{1-2} \cline{4-5} 
$T$                & 24             &  & $\beta$            & 0.99           \\ \cline{1-2} \cline{4-5} 
\rev{$N$}                & \rev{20}             &  & \rev{$\epsilon$}           & \rev{0.2}           \\ \cline{1-2} \cline{4-5} 
\begin{tabular}[c]{@{}l@{}}$E_{begin}$,\\ $E_{target}$\end{tabular} &
  \begin{tabular}[c]{@{}l@{}}Scarce: 16 J \\ \rev{Abundant: 144 J}\end{tabular} &
   &
  \begin{tabular}[c]{@{}l@{}}EH standard\\  deviation\end{tabular} &
  \begin{tabular}[c]{@{}l@{}}0\%, 5\%, 10\%,\\ 15\%, 20\%, 25\%\end{tabular} \\
  \bottomrule
\end{tabular}
\vspace{-5mm}
\end{table}

\rev{\noindent \textbf{Choice of energy storage:} Supercapacitors are attractive alternatives to batteries due to their more efficient charging. 
In our prototype device we used a lithium-ion battery since it can retain its charge for a longer period of time and has less leakage. 
With choice, the ECO framework is built around a battery powered system. 
However, this is not a stringent requirement, ECO can also work with supercapacitors.}

\rev{\noindent \textbf{Cost and comfort of the wearable prototype:} The MFC8528P2 piezoelectric element costs 165\$ each and the SP3-37 PV-Cell costs 3\$ each. 
In addition, the bill of materials of the prototype board is currently less than 20\$. 
Our driver applications are in the medical context, such as falling prevention/detection in elderly care, or freezing of gait detection in Parkinson’s disease patients. 
To this end, we place our device on a comfortable knee sleeve, where the PV-cells are placed on the sleeve and piezoelectric elements are inserted into pockets on the sleeve. 
Our experience with this setup shows that the sleeve is comfortable and wearing it does not impact people’s behavior in any way.
}

\vspace{-2mm}
\subsection{Implementation Overhead}
\label{sec:expresults}

The execution times and energy consumption of the ECO framework and the prior approach~\cite{bhat2017near} are measured on the prototype.
We also execute the iterative gradient projection algorithm (Algorithm~\ref{alg:iter}) at every interval $t$ 
to obtain a reference utility and measured its energy consumption
(tolerance: 1$\times$10$^{-6}$, $stepSize$=1$\times$10$^{-3}$).

\renewcommand{\arraystretch}{0.6}
\begin{table}[!b]
\vspace{-4mm}
\centering
\caption{Execution time and energy measurements}
\vspace{-2mm}
\label{tab:time_power}
\begin{tabular}{@{}cllll@{}}
\toprule
\multicolumn{1}{l}{}                                                            &         & \textbf{Iterative}	& \textbf{ECO} 			& \multicolumn{1}{l}{\textbf{{[}1{]}}} \\ \midrule
\multirow{2}{*}{\textbf{\begin{tabular}[c]{@{}c@{}}Time\\ (s)\end{tabular}}}    & once    & N/A               	& 9.93$\times10^{-1}$   & N/A       \\
                                                                                & every t & 7.03$\times10^{1}$	& 2.50$\times10^{-2}$   & 2.00$\times10^{-3}$   \\ \midrule
\multirow{2}{*}{\textbf{\begin{tabular}[c]{@{}c@{}}Energy\\ (mJ)\end{tabular}}} & once    & N/A               	& 3.48           		& N/A       \\
                                                                                & every t & 2.46$\times10^{2}$	& 8.75$\times10^{-2}$   & 7.00$\times10^{-3}$   \\ \midrule
\multicolumn{2}{l}{\textbf{Normalized Utility$^*$}}                                           & 1.0               	& 0.98             		& 0.83    \\ \bottomrule
\end{tabular}
\\ 
\vspace{1mm}
\raggedright
\textbf{$^*$}The utility is normalized with respect to the iterative technique.
\end{table}

Table~\ref{tab:time_power} shows the execution time, energy consumption, and utility for the three approaches for a typical user.
ECO achieves nearly the same utility as the iterative algorithm while having significantly less execution time and energy consumption.
\rev{Specifically, the energy overhead of ECO is only 87.5~$\mu$J/hour, which is negligible compared to the energy harvested in an hour.}
The initial run of the iterative algorithm in ECO uses larger
tolerance and step size (1$\times$10$^{-3}$, 5$\times$10$^{-3}$), 
since the rollout phase makes up for the loss in optimality. 
The execution time of the correction in each interval is only 25 ms. 
The approach in~\cite{bhat2017near} takes virtually zero time to execute since it is a closed-form solution.
However, it suffers from low utility, which renders it impractical.
These results show that the ECO framework achieves the same utility as the iterative algorithm while having 1000$\times$ smaller energy overhead.

\rev{\noindent \textbf{Complexity:} The ECO framework runs the iterative gradient projection algorithm (Algorithm 1) once to compute initial allocations. 
Then, in the rollout phase, it iterates over $N$ branches (O($N$)) to try different coefficients for the current interval, as depicted in Figure~\ref{fig:rollout}. 
After each branch is visited and a total utility for each is computed, these utility values are sorted ($N$log$N$) and checked for constraint violations starting from the highest. 
Therefore, the complexity of the proposed ECO framework is O($N$) + $N$log($N$).}

\begin{figure*}[!b]
\footnotesize
\begin{tabular}{cccc}
& \textbf{Energy Allocations} & \textbf{Battery Energy} & \textbf{Energy Harvesting} \\  
\rotatebox{90}{\parbox[c]{4cm}{\centering \textbf{High Std. Dev.,\\Abundant Energy}}} &
\includegraphics[width=0.3\textwidth]{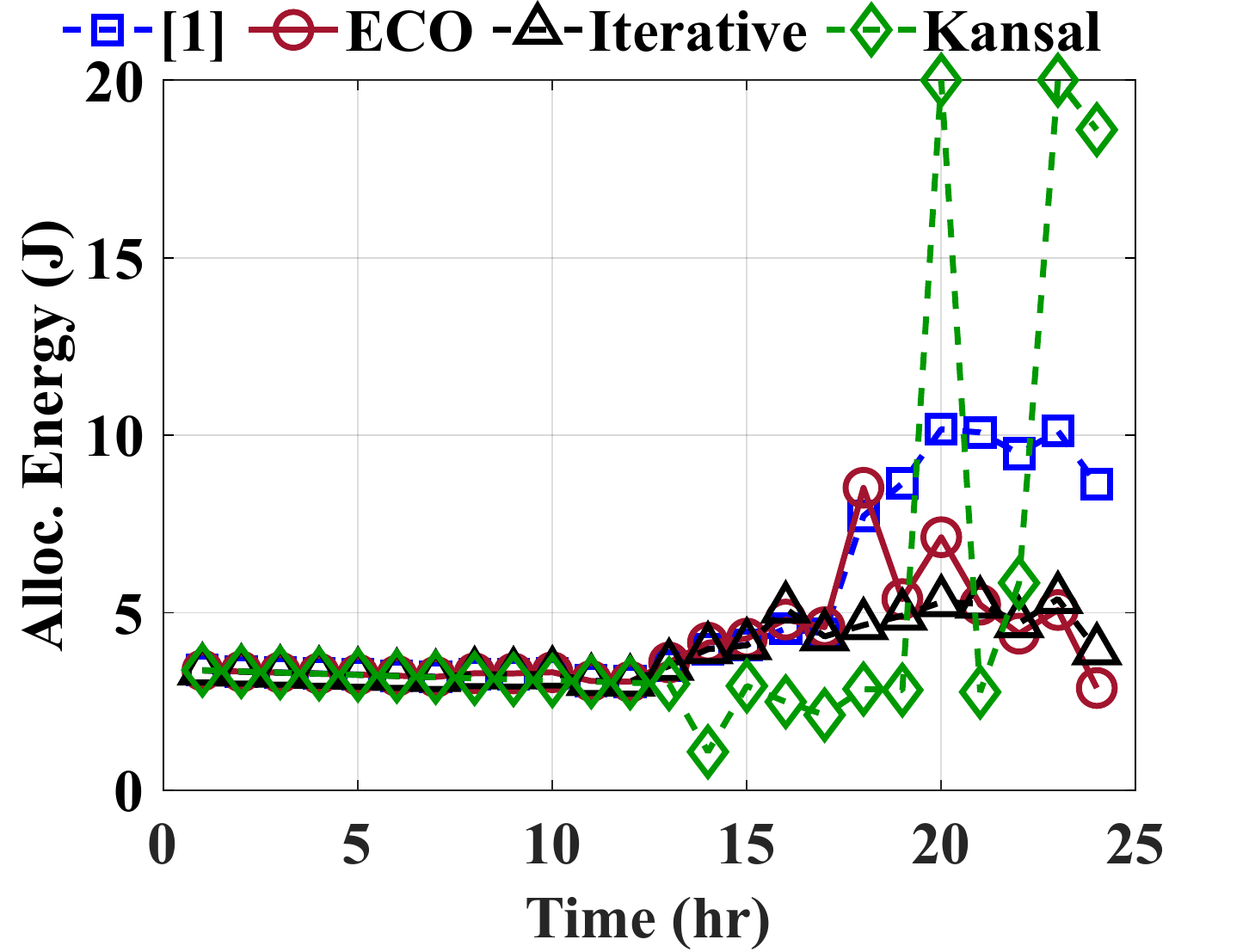} &   \includegraphics[width=0.3\textwidth]{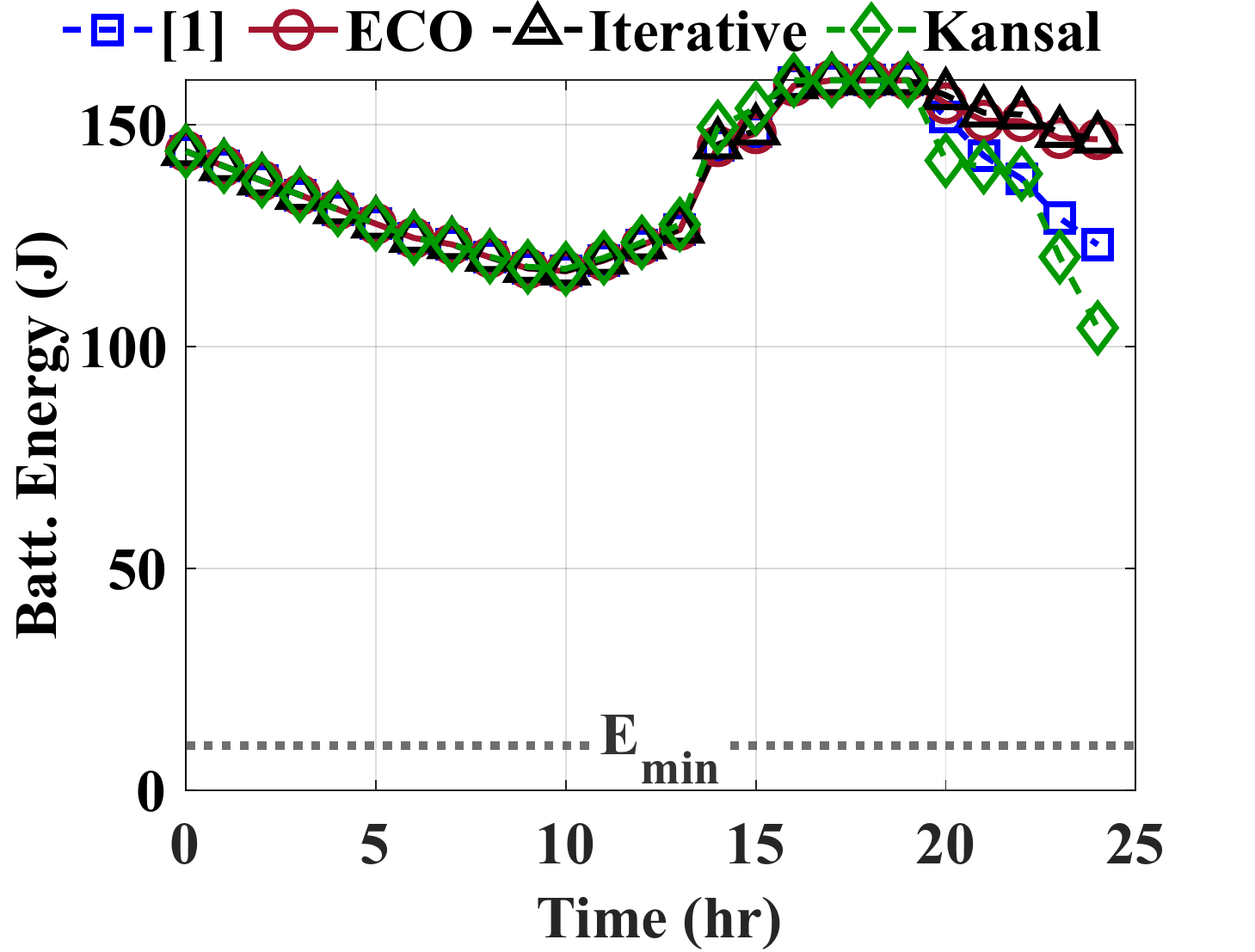} & \includegraphics[width=0.3\textwidth]{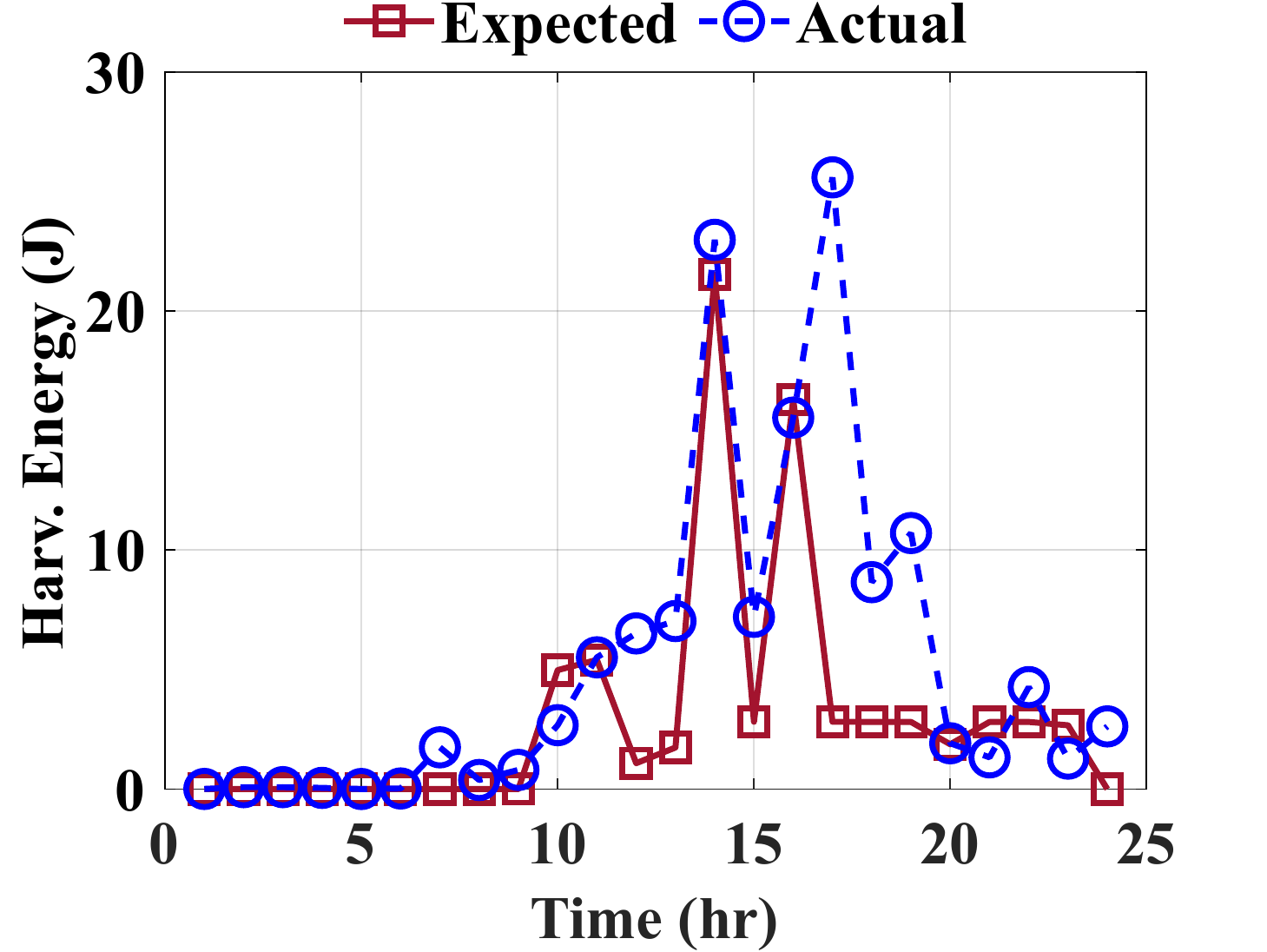}\\
& (1a) & (1b) & (1c)\\[6pt]
\rotatebox{90}{\parbox[c]{4cm}{\centering \textbf{Low Std. Dev.,\\Scarce Energy}}} & 
\includegraphics[width=0.3\textwidth]{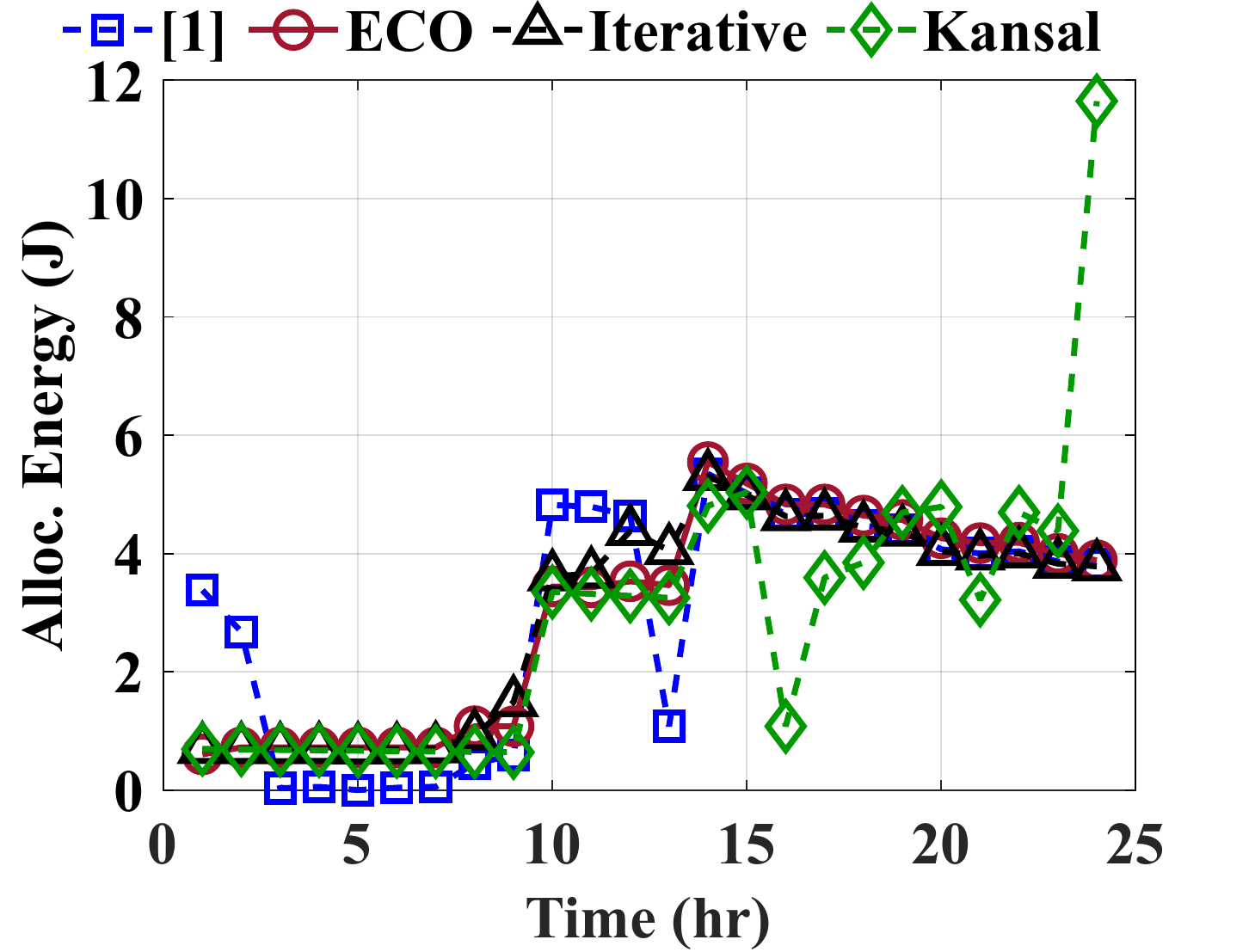} &   \includegraphics[width=0.3\textwidth]{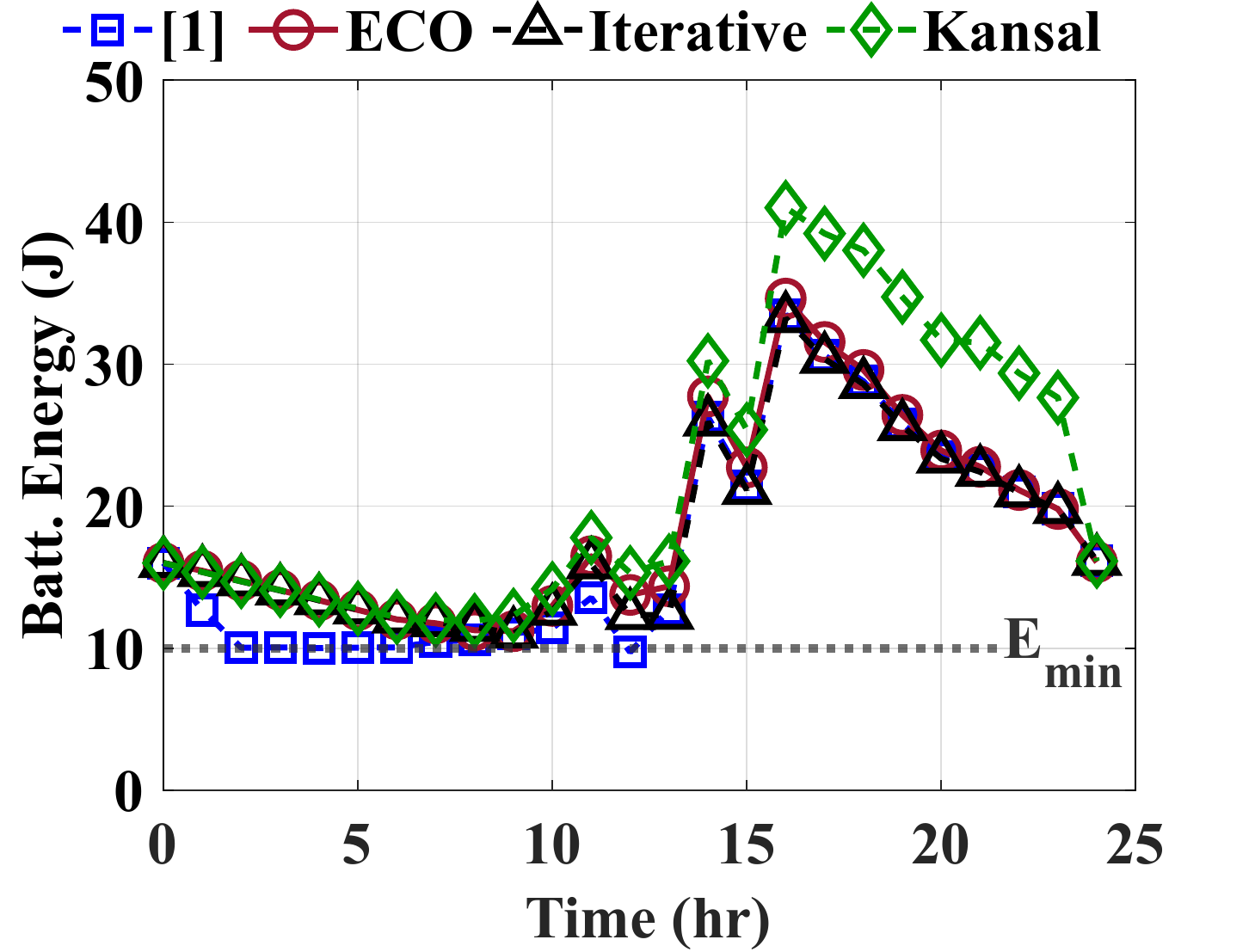} & \includegraphics[width=0.3\textwidth]{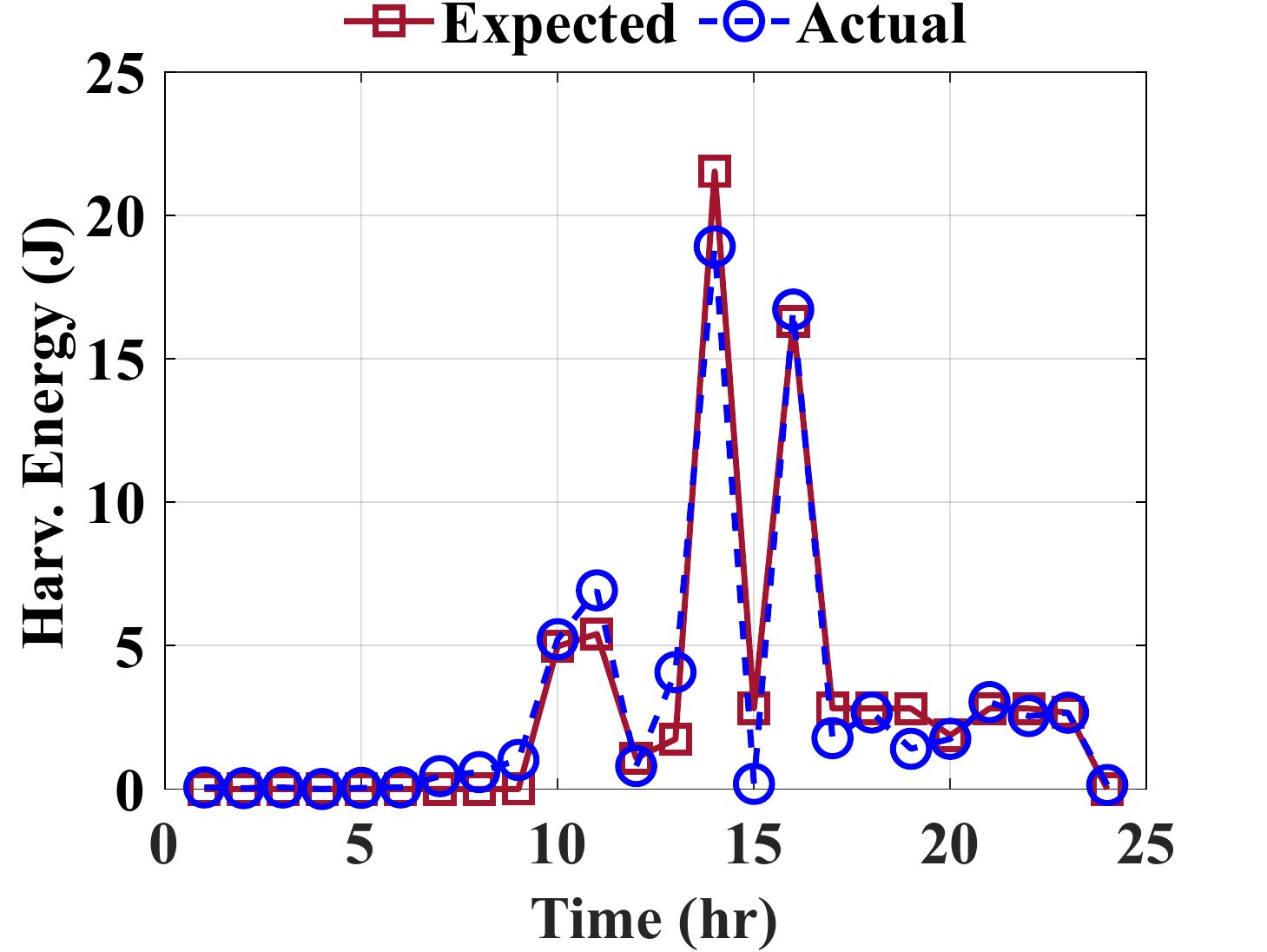}\\
& (2a) & (2b) & (2c) \\[6pt]
\rotatebox{90}{\parbox[c]{4cm}{\centering \textbf{High Std. Dev.,\\Scarce Energy}}} & 
\includegraphics[width=0.3\textwidth]{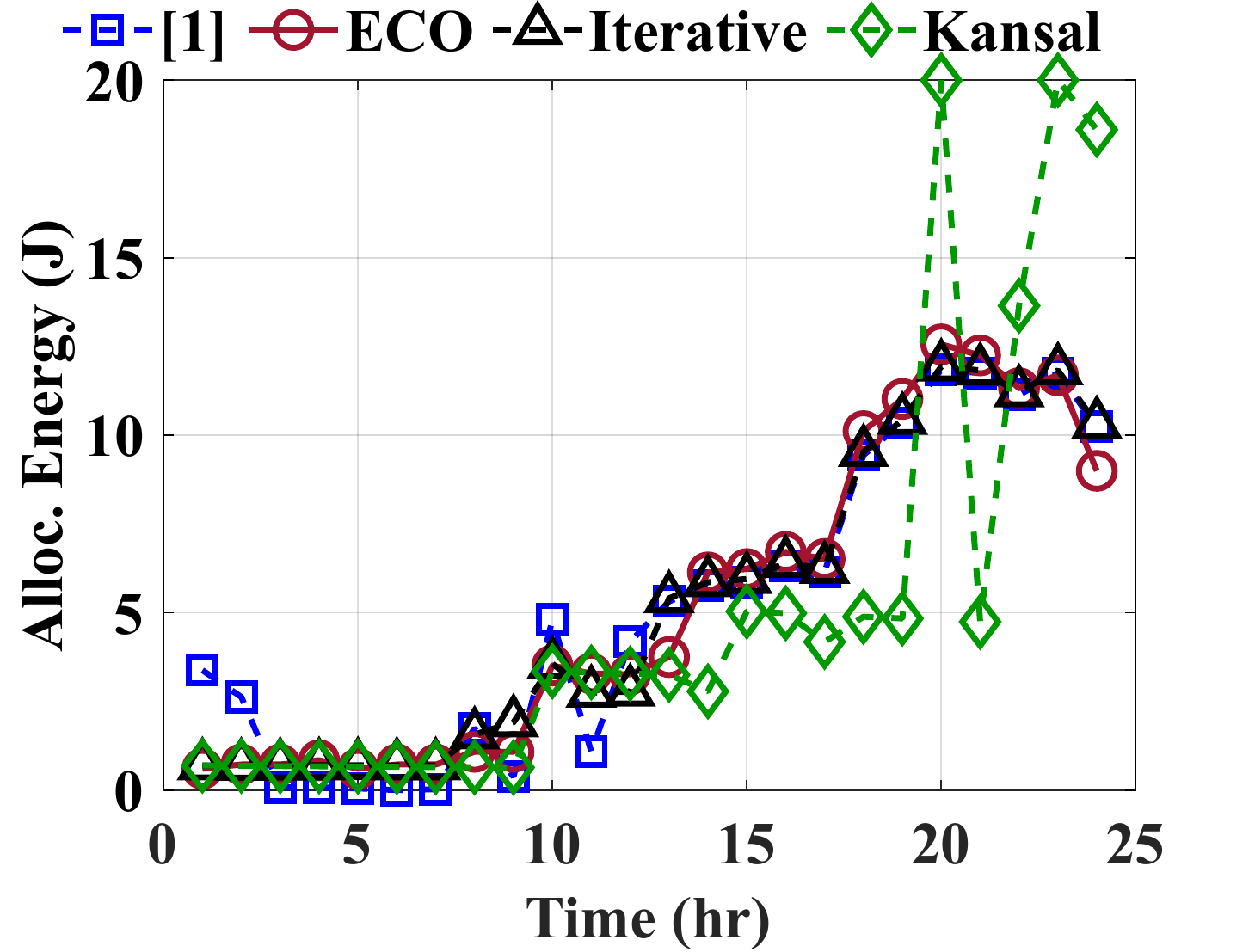} &   \includegraphics[width=0.3\textwidth]{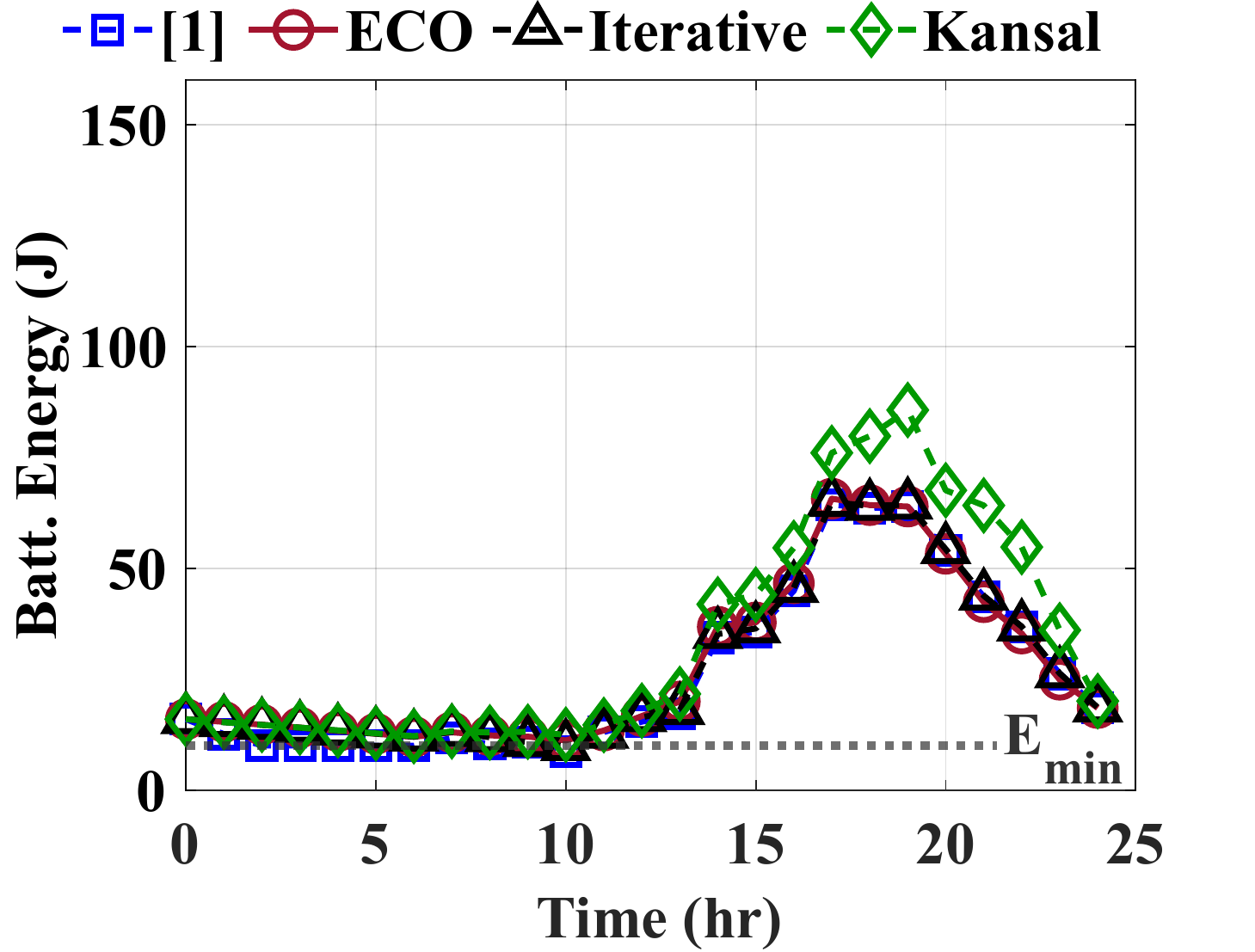} & \includegraphics[width=0.3\textwidth]{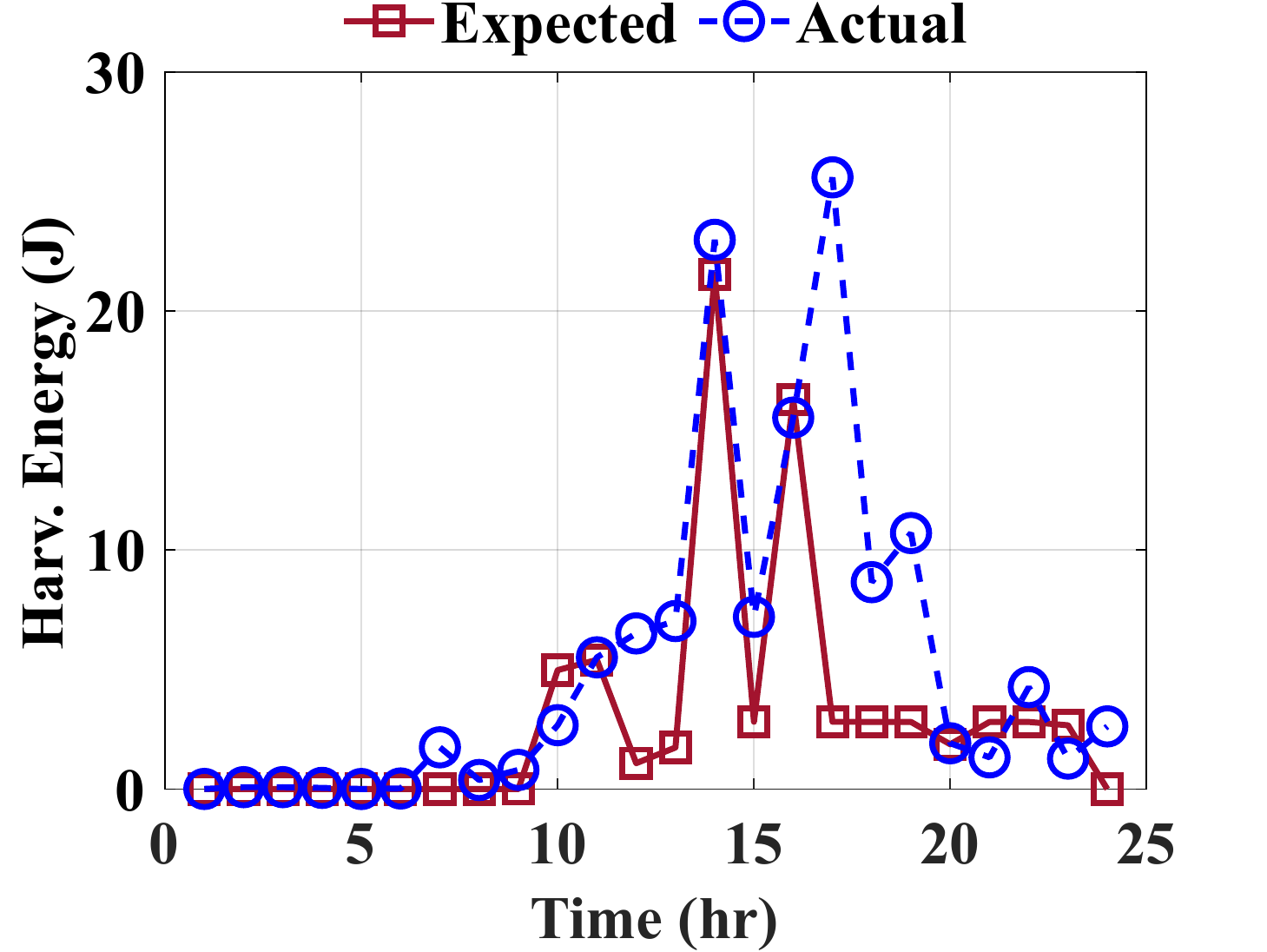}\\
& (3a) & (3b) & (3c) \\[6pt]
\end{tabular}
\vspace{-4mm}
\caption{Comparison of energy allocations and battery energy (b.e.) evolution under different conditions for the same user: 1\textsuperscript{st} row) 25\% standard deviation in EH and 144 J initial b.e., 2\textsuperscript{nd} row) 5\% standard deviation in EH and 16 J initial b.e., 3\textsuperscript{rd} row) 25\% standard deviation in EH and 16 J initial b.e. }
\label{fig:res1}
\normalsize
\end{figure*}

\vspace{-2mm}
\subsection{Daily Energy Allocation Simulation Results}
\label{sec:results}

\rev{This section demonstrates the performance of the ECO framework in multiple steps.
We first compare the daily energy allocations and evolution of battery energy obtained by ECO to those obtained by the iterative approach and prior work in the literature~\cite{bhat2017near, kansal2007power}. 
Moreover, the proposed framework's effectiveness is demonstrated by the increase in average utility over the prior approach across all 4772 users in the dataset. 
We also apply ECO to a non-logarithmic utility function to demonstrate that it can support arbitrary functions. 
} 

\subsubsection{Evaluations using logarithmic utility function}
\label{sec:results_logutility}

\begin{figure*}[!b]
\centering
\vspace{-2mm}
\begin{minipage}{0.95\linewidth}
  \centering
 \rotatebox{90}{\parbox[c]{3cm}{\centering \textbf{\small{~~~~~Abundant Energy}}}}
\includegraphics[width=0.3\linewidth]{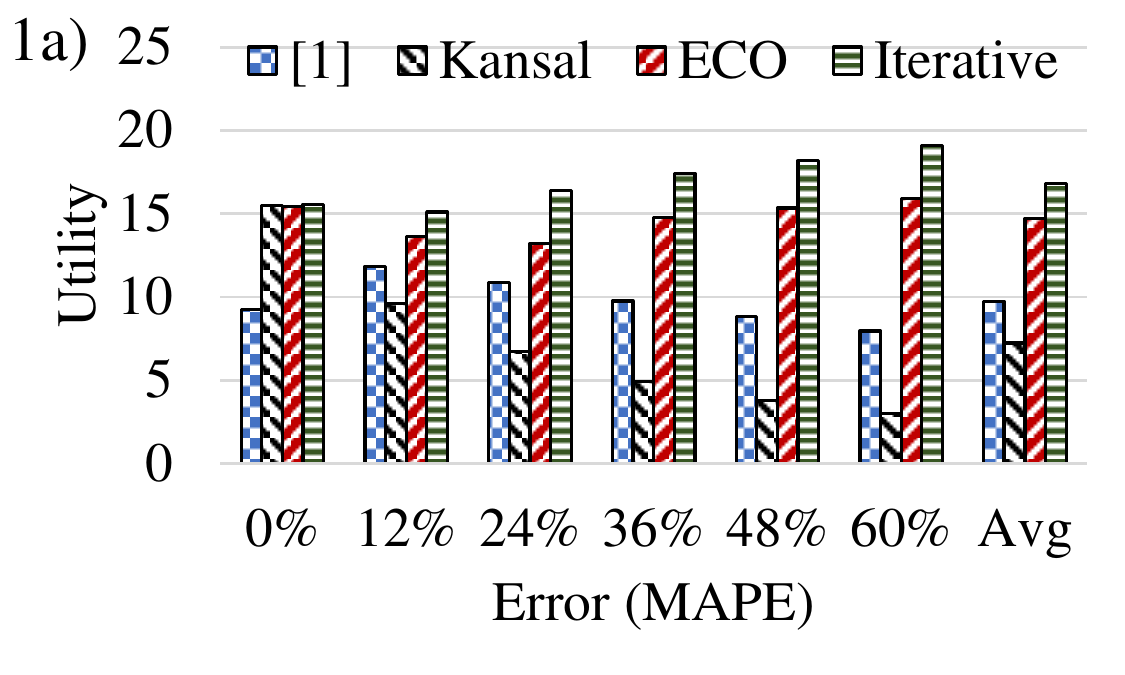}
\hspace{3mm}
\includegraphics[width=0.3\linewidth]{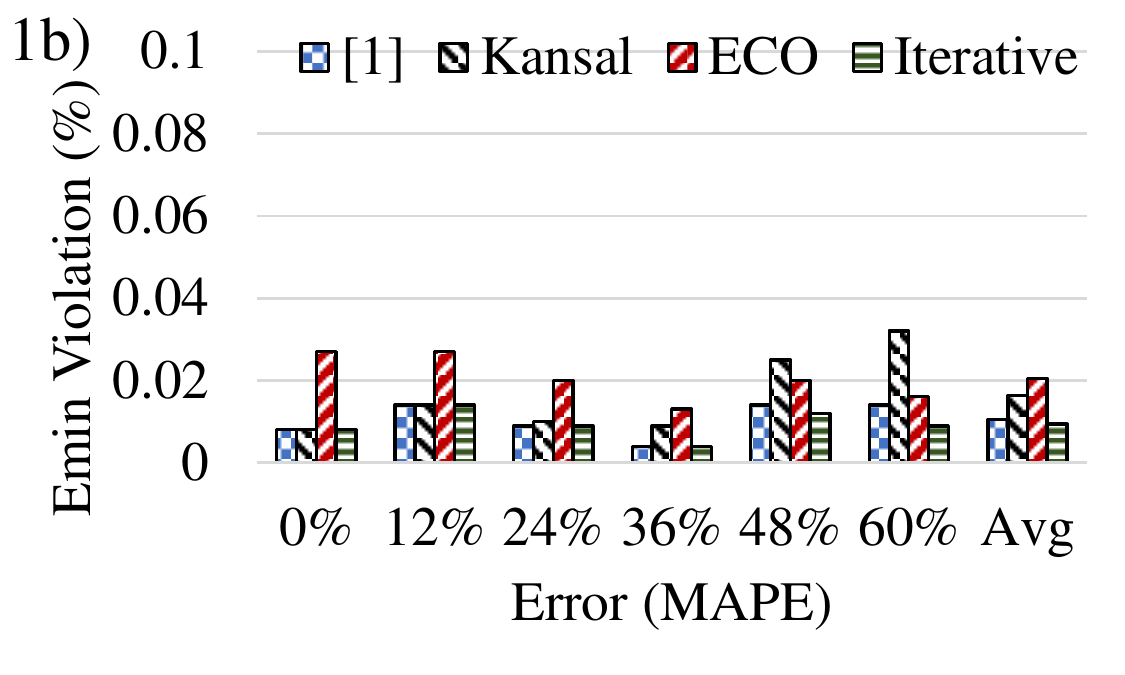}
\hspace{3mm}
\includegraphics[width=0.3\linewidth]{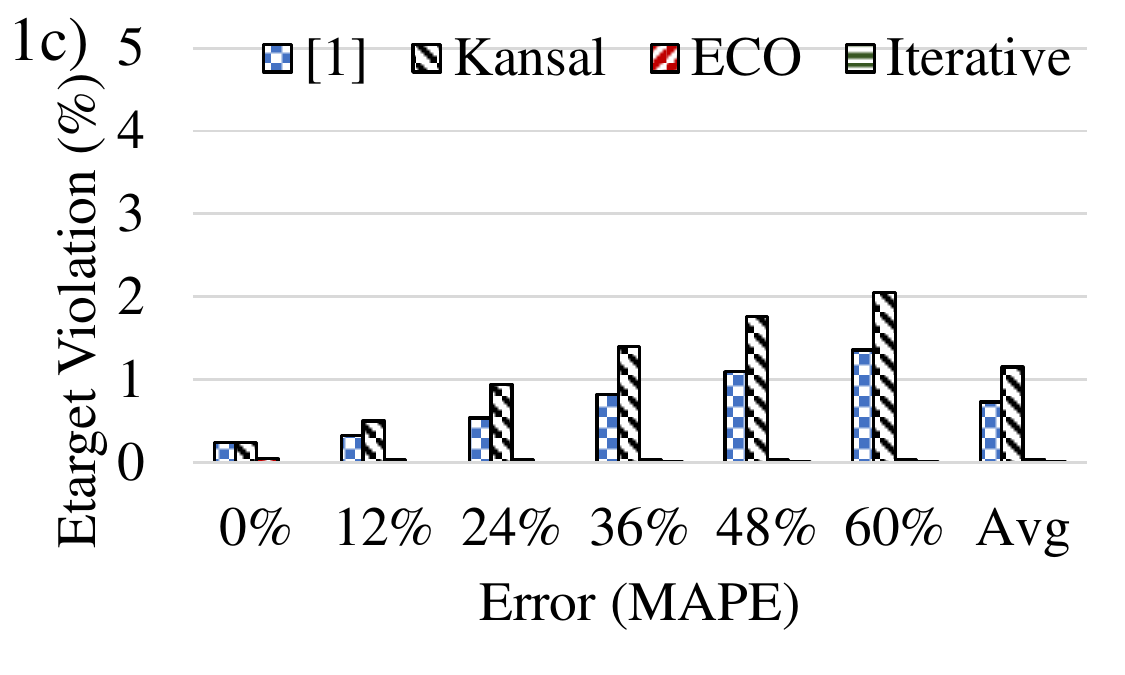}
\end{minipage}
  \vspace{-5mm}
  \begin{minipage}{0.95\linewidth}
    \centering
 \rotatebox{90}{\parbox[c]{3cm}{\centering \textbf{~~~~~\small{Scarce Energy}}}}
\includegraphics[width=0.3\textwidth]{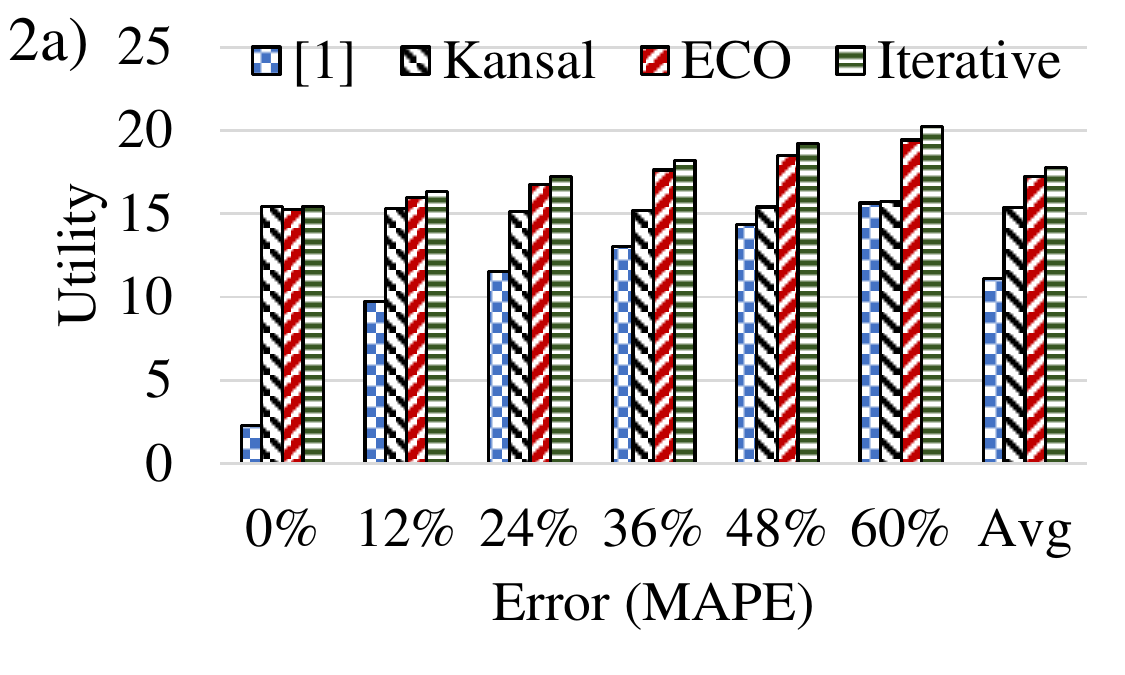}  
\hspace{3mm}
\includegraphics[width=0.3\textwidth]{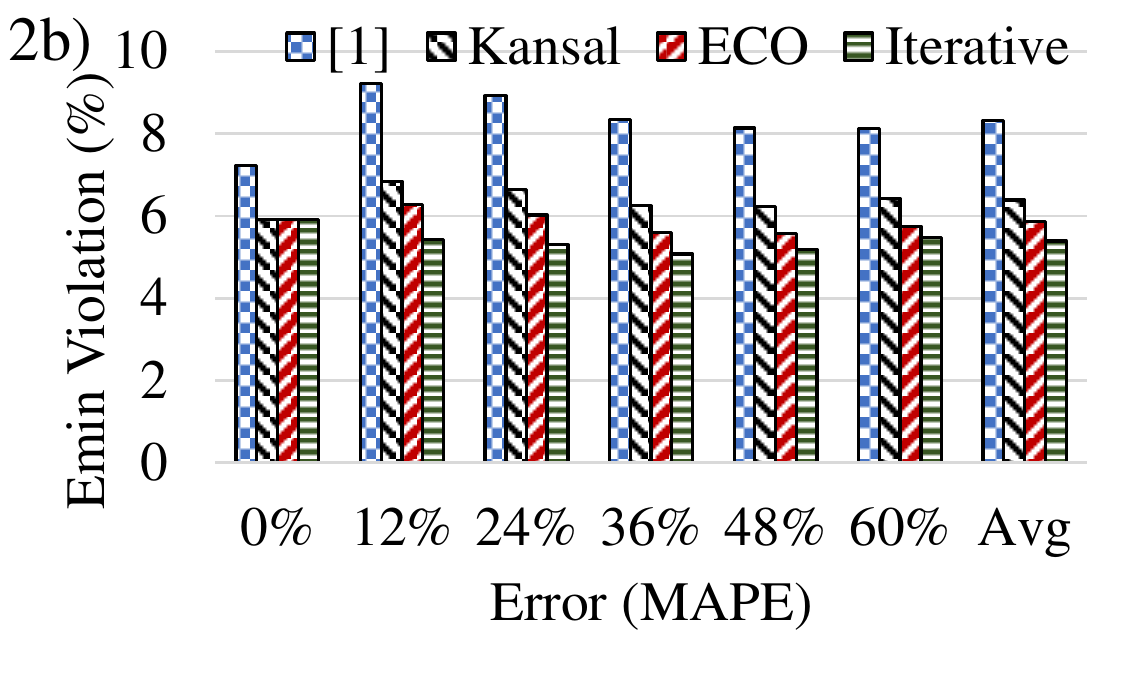}
\hspace{3mm}
\includegraphics[width=0.3\textwidth]{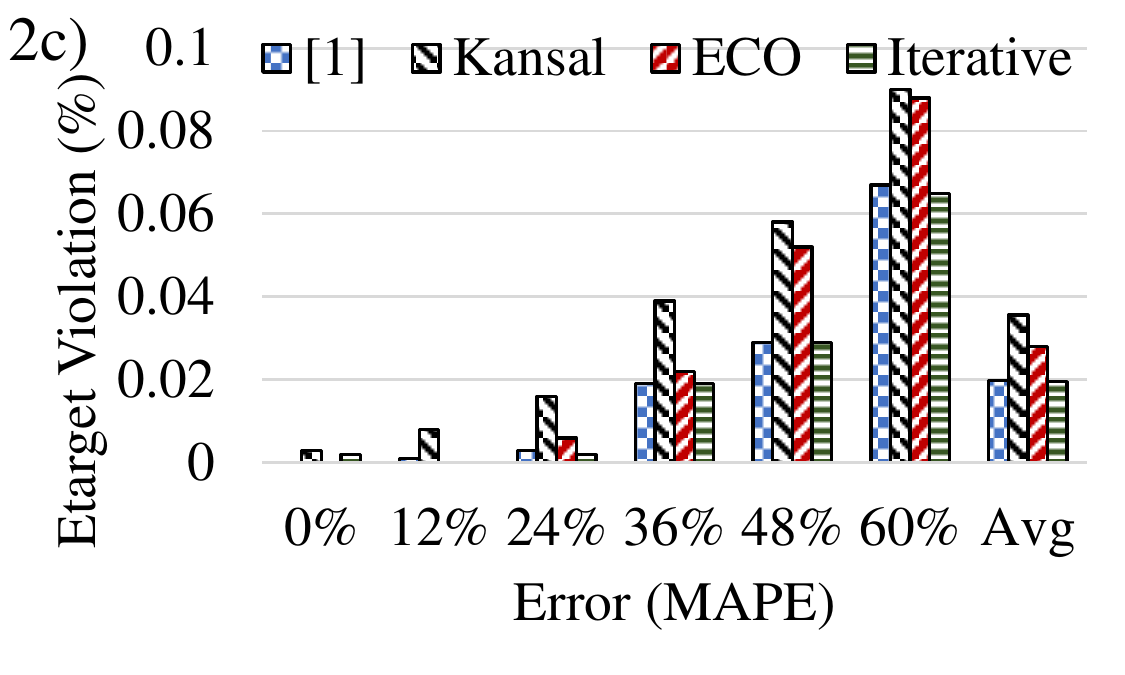}
\end{minipage}
\vspace*{2mm}
\caption{\rev{Comparison of Utility, $E_{min}$ violation and $E_{target}$ violation occurrences: 1\textsuperscript{st} row) 144 J initial battery energy, 2\textsuperscript{nd} row) 16 J initial battery energy.}}
\label{fig:res2}
\end{figure*}

We first use the logarithmic utility function defined in Section~\ref{sec:overview} to demonstrate the performance of ECO.
To this end, we first use Algorithm~\ref{alg:iter} at each interval to recompute the optimal energy allocations as the day progresses.
Second, we use two prior approaches in the literature~\cite{bhat2017near, kansal2007power}.
Finally, we run ECO and compare the obtained energy allocation and battery energy curves with the other solutions.

\noindent \textbf{Energy allocations and battery energy throughout the day:}
\rev{Figure~\ref{fig:res1} compiles the results for a typical user
, who is chosen randomly among the 4772 users in the dataset.
The columns in the figure show the hourly energy allocations, battery energy and EH values throughout the day, respectively.
To highlight the shortcomings and advantages of either approach, we focus on three days with different battery energy and EH conditions.
The rows in Figure~\ref{fig:res1} correspond to these different days. 
Specifically, the first row corresponds to the case with abundant initial battery energy and high standard deviation in energy harvesting as shown in Figure~\ref{fig:res1}-1c. 
In this case, the relaxed $E_{min}$ constraint is not triggered from Figure~\ref{fig:res1}-1b, due to the abundant starting energy at the beginning of the day.
However, the relaxed $E_{max}$ constraint is triggered around 4PM which causes the energy allocations in the later half of the day deviate from the optimal, as shown in Figure~\ref{fig:res1}-1a.
The deviation from the optimal is much smaller for the ECO framework.
Moreover, prior approaches violate the $E_{target}$ constraint at the end of the day, whereas ECO satisfies it.}

\rev{The second row in Figure~\ref{fig:res1} corresponds to scarce initial battery energy and low standard deviation in EH.
In this case, the approach in~\cite{bhat2017near} over-allocates energy at the beginning of the day and triggers the minimum energy constraint only a couple of hours into the day due to the amount of energy expected to be harvested later.
The work in~\cite{kansal2007power} under-allocates energy later in the day and deviates from the optimal battery level curve as can be seen in Figure~\ref{fig:res1}-2b.
In contrast, ECO behaves more conservatively and follows the energy allocation and battery energy curves obtained by the iterative approach much closer, as depicted in Figures~\ref{fig:res1}-2a and 2b, respectively.
As a result, the ECO framework obtains higher utility than the prior approaches.}

\rev{The third row in Figure~\ref{fig:res1} corresponds to the scenario with scarce initial battery energy and high standard deviation in EH.
Similar to the previous case, the prior approaches under- or over-allocate the energy, causing battery energy to deviate from the optimal.
The proposed ECO framework behaves much more conservatively and, as a result, closely follows the iterative solution, as shown in Figures~\ref{fig:res1}-3a and 3b.}

\rev{In general, ECO behaves much closer to the optimal than prior work regardless of the battery level and uncertainty in expected harvested energy. 
This capability of ECO is enabled by the rollout phase, which avoids the piecewise nonlinearity by being less aggressive. 
In contrast, the prior approaches under- or over-allocate the energy and trigger the nonlinearity that is enforced at runtime. }

\noindent \textbf{Utility and $E_{min}$ violations throughout the day:}
\rev{Next, we compare the proposed ECO framework's performance to the prior work in the literature across all 4772 users. 
For this, we analyze the average utility and the $E_{min}$ and the $E_{target}$ violation occurrences for the different approaches.
First, the EH model generates seven separate days per user for each standard deviation value.
For each day, we obtain the results from the ECO framework and the prior work as well as the iterative solution.
Then, the average utility of each approach is found across each day and user.
For example, for the day shown in the 3\textsuperscript{rd} row in Figure~\ref{fig:res1}, ECO achieves 0.91 normalized utility, whereas the prior work in~\cite{bhat2017near} achieves 0.31, and the one in~\cite{kansal2007power} achieves 0.59.
This behavior occurs because the prior approaches under-allocate the energy which is punishing in the logarithmic utility function.  
To obtain the percentage of minimum energy violations, we count the number of intervals the lower threshold in piecewise nonlinearity in Equations~\ref{eqn:Ec_update} and~\ref{eqn:Ec_rollout} are triggered and divide them to total number of intervals (i.e. 24$\times$7 intervals per user).
Similarly, we count the number of target energy violations and divide it to the total number of intervals to obtain the percentage of target energy violations.
These steps are done twice, for scarce and abundant battery energy conditions respectively.}

\rev{Figure~\ref{fig:res2} compares the utility figures in the first column, the $E_{min}$ violation occurrences in the second column and the $E_{target}$ violation occurrences in the third column for all three approaches, as a function of mean absolute percentage error (MAPE) in the expected harvested energy.
The rows correspond to abundant and scarce battery energy conditions, respectively.
In both cases, the proposed ECO framework outperforms the prior approaches as depicted in Figure~\ref{fig:res2}-1a and 2a.
When the battery energy is abundant, all approaches have very small number of $E_{min}$ violations as shown in Figure~\ref{fig:res2}-1b, as expected.
In addition, ECO causes close to zero $E_{target}$ violations, whereas the previous approaches cause significantly more, as depicted in Figure~\ref{fig:res2}-1c.
Under scarce battery energy conditions, the proposed ECO framework again outperforms the prior approaches as illustrated in Figure~\ref{fig:res2}-2a.
Specifically, it obtains between 10\% -- 35\% more utility on the average.
\rev{Moreover, Figure~\ref{fig:res2}-2b shows that $E_{min}$ violations are inevitable for about 5\% of the entire dataset when the battery is scarce even for the iterative algorithm.
When a minimum energy violation happens, the device goes to a sleep state.
The cost of this action is zero utility during sleep periods, which is reflected in the reported utility.
The prior approaches lead to significantly higher $E_{min}$ violations than ECO.
Specifically, ECO achieves close to 3\% lower $E_{min}$ violations than the prior work.}
Finally, all approaches have very small number of $E_{target}$ violations under scarce battery energy as shown in Figure~\ref{fig:res2}-2c.}
\revv{It can be seen that the utility obtained by the optimal iterative algorithm and the other approaches increase as the MAPE increases.
This observation is due to the rectified gaussian distribution forcing the harvested energy to be positive.
It results in higher mean harvested energy over a day as the MAPE increases.
As a result of higher harvested energy, all approaches obtain higher average utility.}

\begin{figure}[t]
\centering
\begin{tabular}{c}
\includegraphics[width=0.3\textwidth]{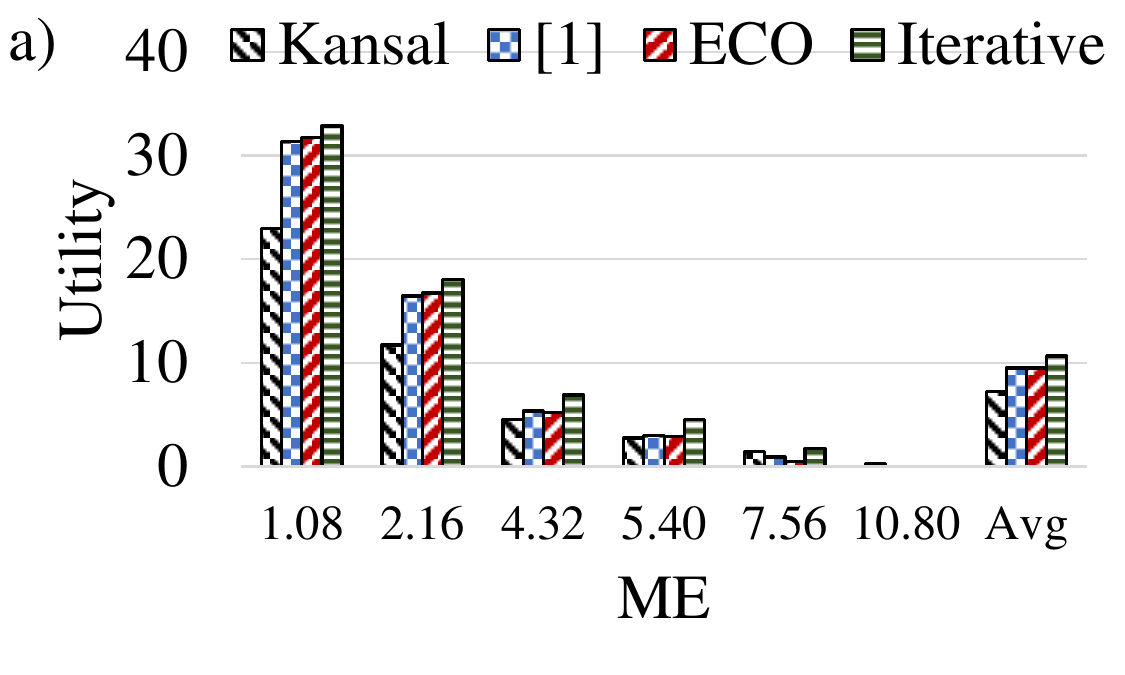} \\
\vspace{-3mm}
\includegraphics[width=0.3\textwidth]{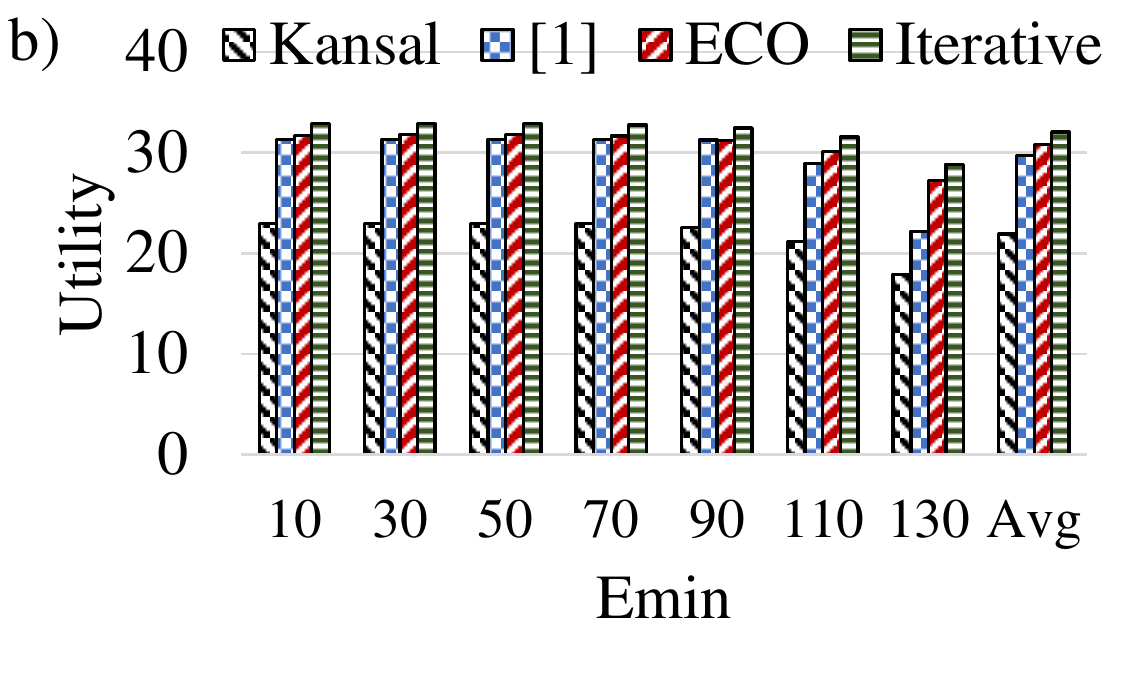} \\
\vspace{-3mm}
\includegraphics[width=0.3\textwidth]{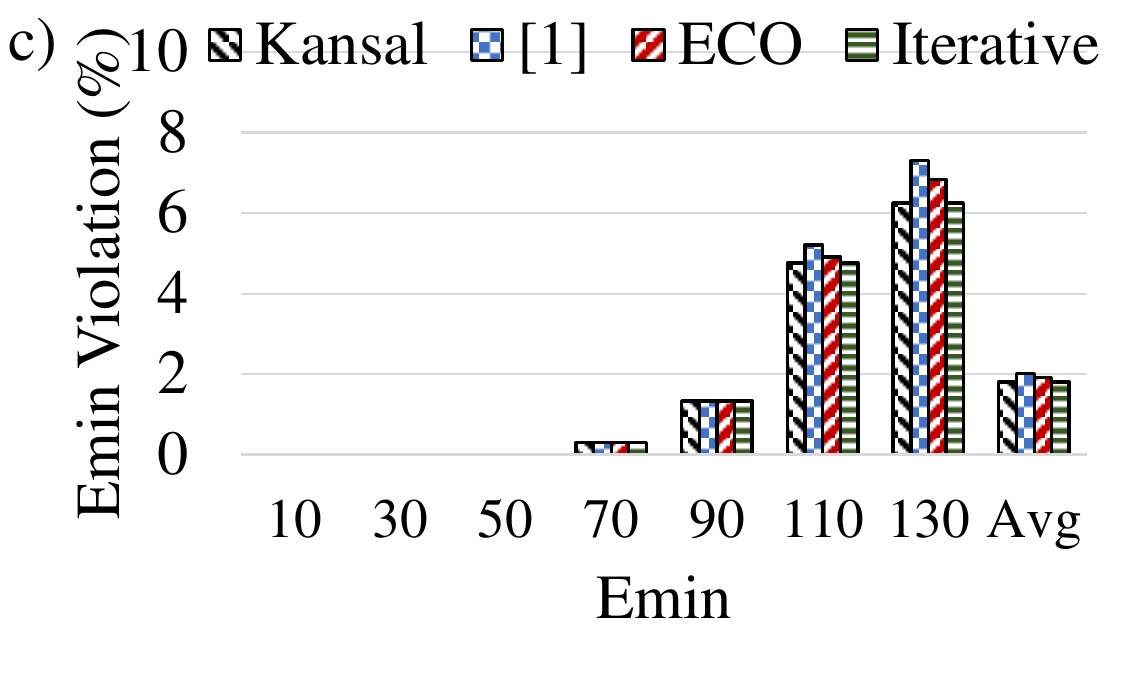} \\
\end{tabular}
\caption{\rev{The effect of a) increasing ME on utility, b) increasing $E_{min}$ on utility, c) increasing $E_{min}$ on $E_{min}$ Violation Percentage}}
\label{fig:res4}
\vspace{-5mm}
\end{figure}

\rev{\noindent \textbf{The effect of $M_E$ and $E_{min}$:} We selected the values for parameters $M_E$ and $E_{min}$ from our prototype as explained in the beginning of Section~\ref{sec:eval}. 
To observe their effects on ECO and the other approaches, we swept the values of $E_{min}$ and $M_E$ independently of each other, as depicted in Figure~\ref{fig:res4}. 
According to these results, increasing $M_E$ has a negative effect on the utility, as shown in Figure~\ref{fig:res4}-a. 
This behavior is expected because $M_E$ is the minimum energy required for obtaining positive utility.
Increasing $M_E$ does not affect the percentage of $E_{min}$ and $E_{target}$ constraint violation occurrences. 
Similarly, increasing $E_{min}$ also degrades the utility, as depicted in Figure~\ref{fig:res4}-b. 
This is because the allocation algorithms are forced to allocate lower energy to prevent violating the $E_{min}$ constraint, which decreases the utility.
In addition, Figure~\ref{fig:res4}-c shows that increasing $E_{min}$ significantly increases the percentage of $E_{min}$ constraint violation occurrences, as expected. 
Increasing $E_{min}$ does not affect the percentage of $E_{target}$ constraint violation occurrences. }

\subsubsection{Evaluations using an arbitrary utility function}
\label{sec:results_logarbitrary}
One of the key advantages of the proposed ECO framework is its general applicability to arbitrary utility functions.
Next, we use the energy vs. accuracy relationship of a gesture recognition application from the literature to demonstrate the performance of the ECO framework with an arbitrary utility function.

\begin{figure}[!t]
\centering
\includegraphics[width=0.85\linewidth]{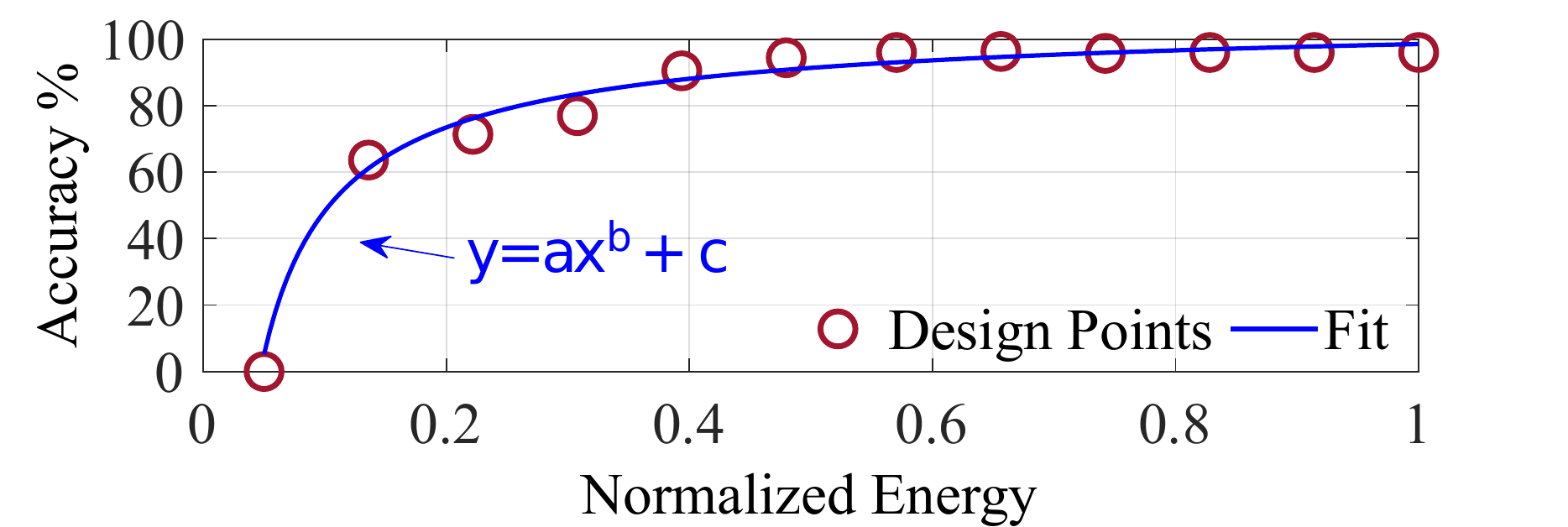}%
\vspace{-3mm}
\caption{Design points for a gesture recognition application and the fitted energy vs. accuracy curve.}
\label{fig:application_fit}
\vspace{-5mm}
\end{figure}

Figure~\ref{fig:application_fit} shows 12 design points for a gesture recognition application from the literature~\cite{park2020energy}.
A function with the form $y=ax^b+c$ fits much better to these data points than the logarithmic function in Section~\ref{sec:overview} (i.e. R$^2$ value for the two functions are 0.96 and 0.82, respectively).
Therefore, one can use the discrete points, or the fit shown with the blue curve in Figure~\ref{fig:application_fit} as the utility function (i.e., the utility is defined as the classification accuracy).

\begin{figure*}[!t]
\centering
\begin{minipage}{0.95\linewidth}
  \centering
 \rotatebox{90}{\parbox[c]{3cm}{\centering \textbf{\small{~~~~~Abundant Energy}}}}
\includegraphics[width=0.3\textwidth]{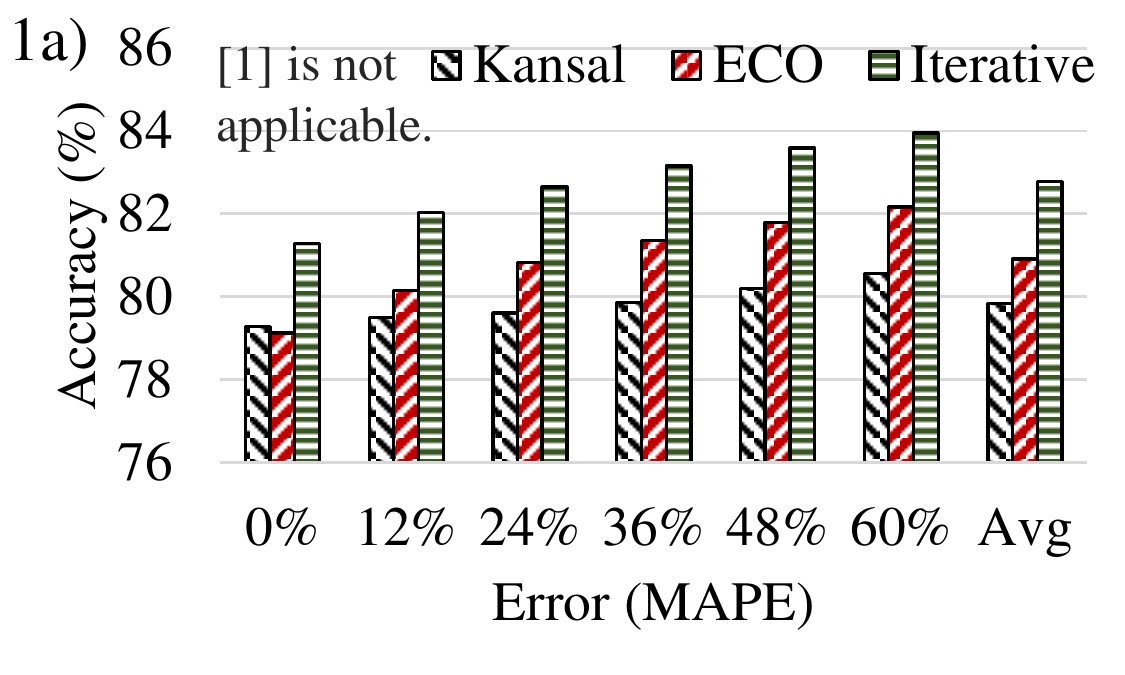}
\hspace{3mm}
\includegraphics[width=0.3\textwidth]{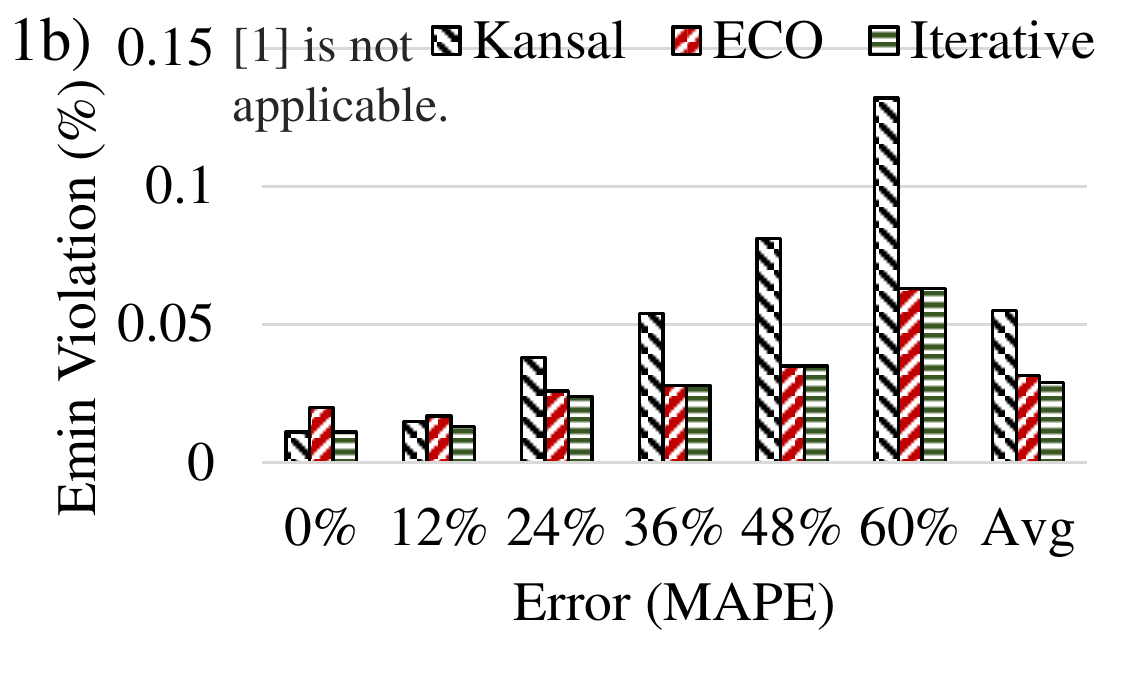}
\hspace{3mm}
\includegraphics[width=0.3\textwidth]{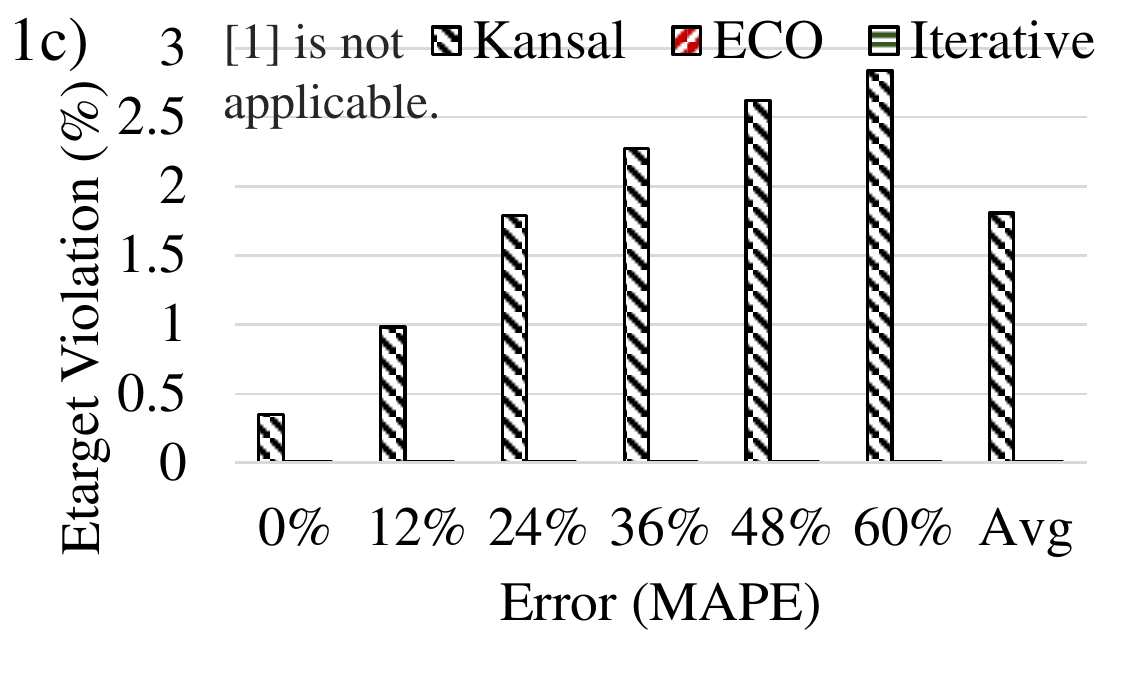}
\end{minipage}
  \vspace{-5mm}
  \begin{minipage}{0.95\linewidth}
    \centering
 \rotatebox{90}{\parbox[c]{3cm}{\centering \textbf{~~~~~\small{Scarce Energy}}}}
\includegraphics[width=0.3\textwidth]{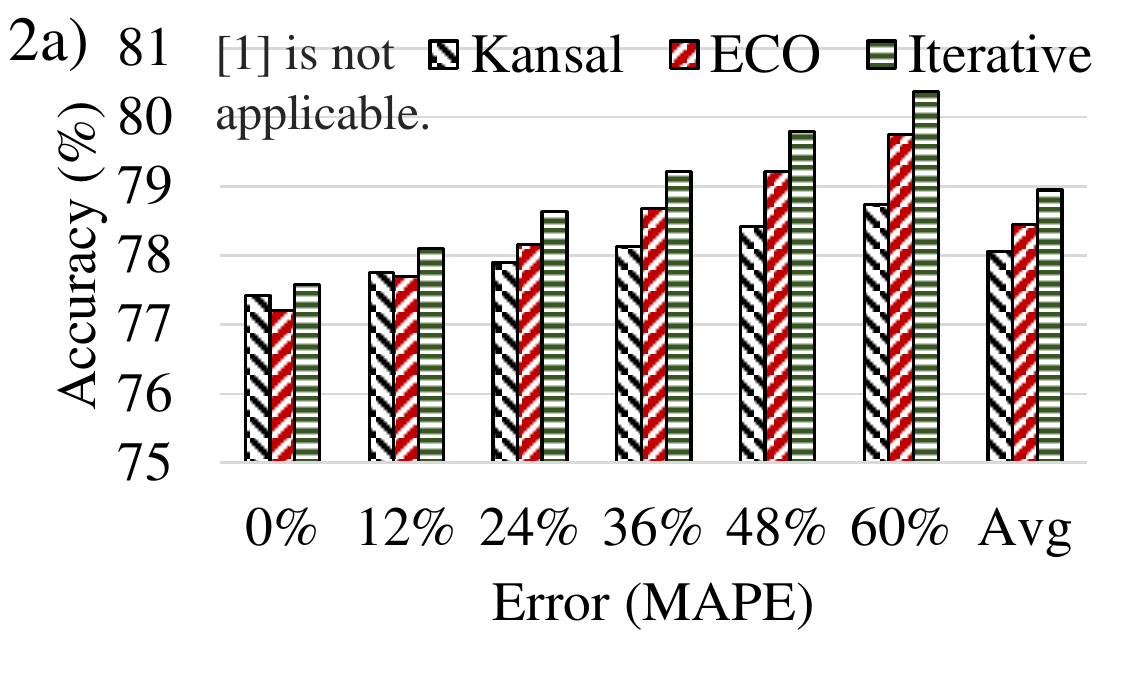} 
\hspace{3mm}
\includegraphics[width=0.3\textwidth]{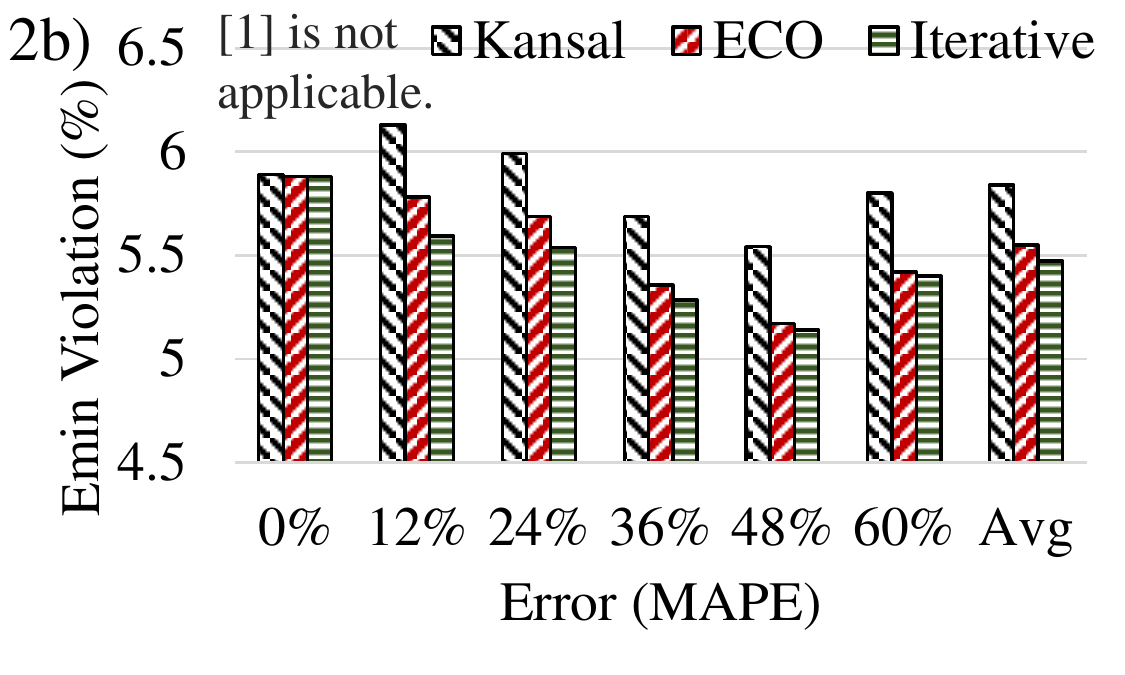}
\hspace{3mm}
\includegraphics[width=0.3\textwidth]{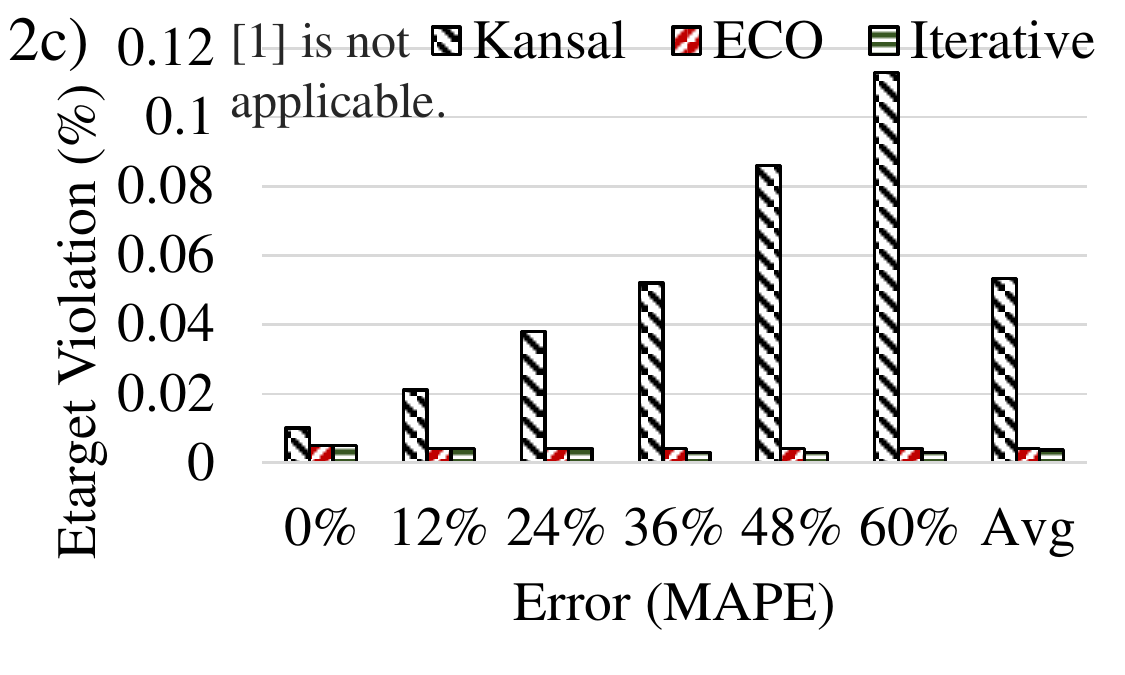}
\end{minipage}
\vspace*{2mm}
\caption{\rev{Accuracy, $E_{min}$ violation and $E_{target}$ violation occurrences:
 1\textsuperscript{st} row) 144 J initial battery energy, 2\textsuperscript{nd} row) 16 J initial battery energy.}}
 \vspace*{-2mm}
\label{fig:res3}
\end{figure*}

\noindent \textbf{Accuracy and $E_{min}$ violations throughout the day:}
\rev{We follow the same procedure used in generating Figure~\ref{fig:res2} to demonstrate the performance of the proposed ECO framework with the utility function depicted in Figure~\ref{fig:application_fit}.
Figure~\ref{fig:res3} shows the accuracy in the first column, $E_{min}$ violations and $E_{target}$ violations across all users in the second and third column, respectively.
The rows correspond to abundant and scarce battery energy conditions.}

\rev{ECO obtains similar accuracy to the iterative solution when the battery energy is abundant and outperforms the approach in~\cite{kansal2007power}, as shown in Figure~\ref{fig:res3}-1a.
More specifically, ECO obtains close to 2\% higher accuracy on the average.
In addition, ECO causes $E_{min}$ violations in less than 0.03\% of the total 4772$\times$7$\times$24 intervals as shown in Figure~\ref{fig:res3}-1b.
In contrast, the prior work causes two times of that.
Moreover, Figure~\ref{fig:res3}-1c illustrates that ECO significantly outperforms the prior work in $E_{target}$ violations. 
ECO causes essentially no violations, whereas the prior work has $E_{target}$ violations in 2\% of the total number of days.
Under scarce battery energy conditions, the performance of all three approaches deteriorate, as expected.
However, ECO can still achieve more accuracy compared to the prior work.
For example, Figure~\ref{fig:res3}-2a shows that at 60\% MAPE in EH, the iterative solution and the ECO framework achieve 80.37\% and 79.75\% accuracy, respectively.
Moreover, Figure~\ref{fig:res3}-2b shows that $E_{min}$ violation occurrences are within less than 1\% of the optimal when battery energy is scarce.
In contrast, the prior work causes more $E_{min}$ violations for all error levels in harvested energy.
Finally, all approaches have very small number of $E_{target}$ violations under scarce battery energy as expected, as shown in Figure~\ref{fig:res3}-2c.}

\subsection{\rev{Three-year Energy Allocation Simulation Results}}
\rev{This section demonstrates the performance of the ECO framework in continuous use throughout three years. 
For this, we have generated an annual energy harvesting profile for the median users of the four clusters which are shown in Figure~\ref{fig:EHMODEL}. 
We used the highest model error (i.e. 25\% standard deviation - 60\% MAPE) when generating each day. 
In addition, our model also incorporates the seasonal solar energy increase during the summer months. 
Note that this only has an impact on the harvested energy if the user spends time outdoors.}

\rev{\noindent \textbf{Battery degradation, leakage, and inefficiency:} Recent studies show 5\% to 10\% annual capacity degradation for lithium-ion batteries~\cite{ecker2012development}. 
Following this, our three-year analysis includes a daily linear degradation that corresponds to 5\% annual degradation. 
Similarly, the work in~\cite{ecker2012development} claims that a battery can complete its lifetime in as short as 2 years if it is operated near its full capacity most of the time. 
If operated at 50\% capacity, its lifetime can be as long as 6 years. 
We argue that the ECO framework can be configured to achieve this long lifetime by setting the $E_{begin}$ and $E_{target}$ values accordingly, such that the battery is kept in the middle region most of the time. 
We do not model the battery leakage separately as it is already considered in the calculation of 1.08 J as $M_E$.
}
\begin{figure}[!t]
\vspace{-4mm}
\centering
\begin{tabular}{c}
\includegraphics[width=0.3\textwidth]{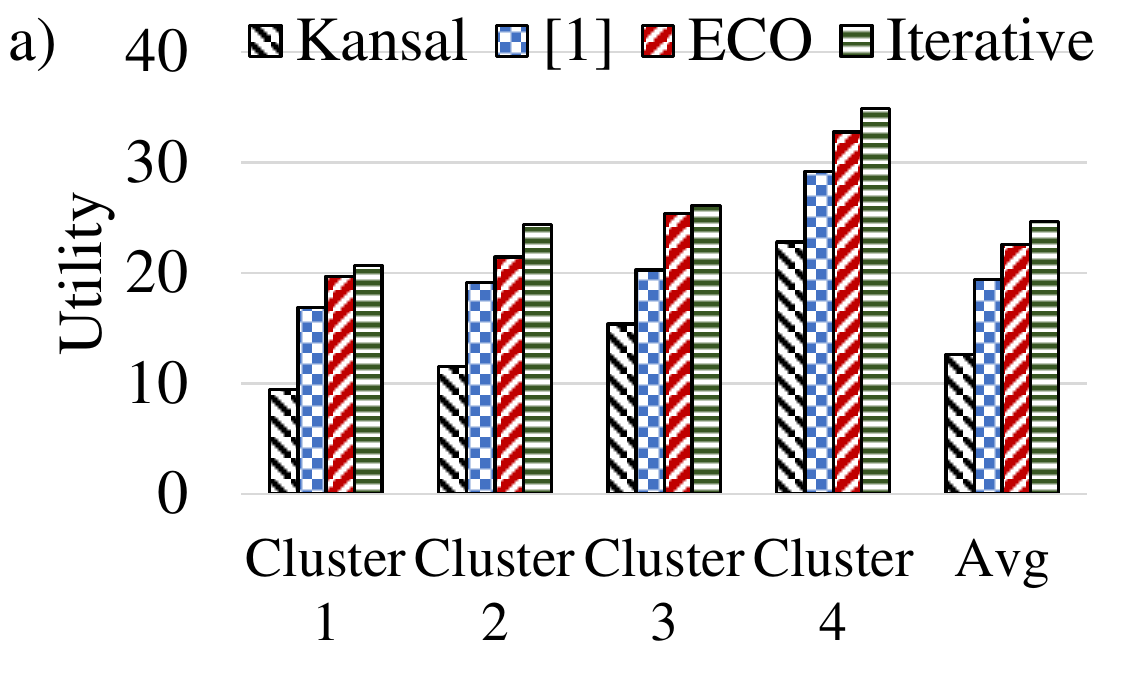} \\
\includegraphics[width=0.3\textwidth]{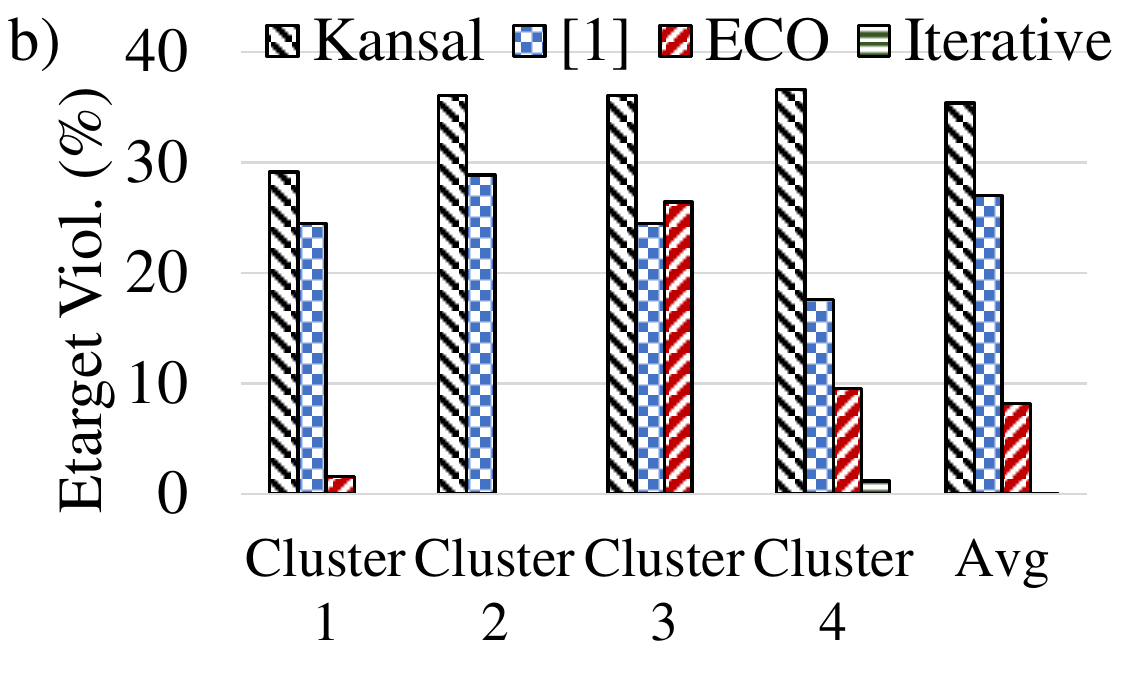} \\
\end{tabular}
\caption{\rev{3-year simulation results. a) Utility b) $E_{target}$ violations (\%)}}
\label{fig:res5}
\vspace{-3mm}
\end{figure}

\rev{
We assume that the device starts with a 90\% battery capacity (i.e. we assume an initial charging of the battery). 
Then, the battery level at the end of the first day is assigned as the $E_{begin}$ for the next day. 
The $E_{target}$ for this day is also set to the same value. 
The time interval is 1-hour, and the simulation continues for 3-years. 
Therefore, in total there are 365$\times$3$\times$24 intervals. 
Figure~\ref{fig:res5} shows the average utility and $E_{target}$ violations for each cluster over these intervals. 
According to these results, ECO achieves higher utility than the previous approaches while doing significantly less $E_{target}$ violations. 
Specifically, ECO achieves 91\% of the utility obtained by the iterative algorithm, while causing $E_{target}$ violations in less than 10\% of the total number of days. 
In contrast, the approach in~\cite{bhat2017near} and~\cite{kansal2007power} achieve 78\% and 51\% of the optimal utility while causing $E_{target}$ violations in about 30\% of the days. 
This suggests that ECO deviates less from the optimal battery level and is more likely to guarantee energy-neutral operation.}

\section{Conclusions \rev{and Future Directions}}
\label{sec:conc}
IoT devices enable novel applications in a wide range of fields, such as wearable health monitoring, environmental monitoring, and robotics.
Despite this potential, their adoption has been hindered by frequent replacement and maintenance of the limited capacity batteries due to small form-factor constraints.
This paper presented ECO, a runtime framework that allocated the energy harvested from ambient sources optimally to these devices.
The proposed ECO framework first uses an efficient iterative algorithm to compute initial energy allocations at the beginning of a day.
Then, it adjusts these initial allocations at uniform intervals to compensate for the deviations from the expected energy harvesting pattern.
Experiments with data from 4772 users show that the ECO framework achieves significantly higher utility than prior approaches.
Moreover, measurements on a wearable device prototype show that ECO has 1000$\times$ smaller energy overhead than iterative approaches with a negligible loss in utility.

\rev{We are planning to expand the capabilities of our prototype device to keep a log of the incoming energy into the battery. 
This extension will enable us to implement the capability of generating user-specific energy harvesting models discussed in Section V. 
This information is necessary for ECO to operate due to the cumulative energy difference ($\Delta_t$) used in the formulation. 
Once we implement these changes, we plan to design an energy harvesting knee sleeve similar to the descriptions in Section V. 
The knee sleeve will incorporate flexible PV-cell(s) and piezoelectric elements to harvest energy during daily use. 
We will place the prototype device on the knee sleeve and run the ECO framework on it. 
With this, we plan to demonstrate an energy-neutral wearable device prototype that can do complex operations, such as gait analysis or activity detection and recognition.}

\appendices

\section{Proof of Lemma~\ref{lem:correction}}
\label{sec:app1}
Suppose that we are at interval $t=\tau$. 
Then, from Equation~\ref{eqn:EC0} the initial allocation $E^A_\tau$ is given by:
\begin{align}
\label{eqn:app1_eq1}
E^A_\tau=\beta^\tau E^A_0 &= \beta^\tau \frac{E^B_0 - E_{target} + \sum_{t=0}^{T-1}E_t^H} {1+\beta+\beta^2+\ldots+\beta^{T-1} } \nonumber \\ 
&= \frac{E^B_0 - E_{target} + \sum_{t=0}^{T-1}E_t^H}{\frac{\beta^T-1}{\beta^\tau (\beta-1)}}
\end{align}
Let $\widehat E^A_t$ denote the recalculated allocations and $\widehat E^H_t$ stand for the actual energy harvesting values.
Following the same pattern in Equation~\ref{eqn:EC0}, we can write $\widehat E^A_\tau$ as follows:
\begin{align}
\label{eqn:app1_eq2}
\widehat E^A_\tau &= \frac{E^B_\tau - E_{target} + \sum_{t=\tau}^{T-1}E_t^H} {1+\beta+\beta^2+\ldots+\beta^{T-\tau-1} } \nonumber \\
&= \frac{ (E^B_0 + \sum^{\tau-1}_{t=0}\widehat E^H_t -\sum^{\tau-1}_{t=0}\widehat E^A_t) - E_{target} + \sum_{t=\tau}^{T-1}E_t^H}{\frac{\beta^T-\beta^\tau}{\beta^\tau (\beta-1)}}
\end{align}
Let $\overline{\Delta}^H_\tau$ correspond to the sum of differences between actual and expected energy harvesting values until $t=\tau$: 
\begin{equation}
\label{eqn:app1_eq3}
\overline{\Delta}^H_\tau = \sum^{\tau-1}_{t=0} \Delta^H_t= \sum^{\tau-1}_{t=0} \big(\widehat E^H_t - E^H_t\big)
\end{equation}
Similarly, let $\overline{\Delta}^A_\tau$ be the sum of differences between initial and corrected energy allocations until $t=\tau$:
\begin{equation}
\label{eqn:app1_eq4}
\overline{\Delta}^A_\tau = \sum^{\tau-1}_{t=0} \Delta^A_t = \sum^{\tau-1}_{t=0}\big(E^A_t - \widehat E^A_t \big)
\end{equation}
With these, Equation~\ref{eqn:app1_eq2} can be written as:
\begin{align}
\label{eqn:app1_eq5}
\widehat E^A_\tau &= \frac{ \splitfrac{(E^B_0 + \sum^{\tau-1}_{t=0}E^H_t -\sum^{\tau-1}_{t=0} E^A_t + \overline{\Delta}^H_\tau + \overline{\Delta}^A_\tau) }{- E_{target} + \sum_{t=\tau}^{T-1}E_t^H} }{\frac{\beta^T-\beta^\tau}{\beta^\tau (\beta-1)}} \end{align}
Further combining $\sum^{\tau-1}_{t=0}E^H_t + \sum_{t=\tau}^{T-1}E_t^H$ as $\sum_{t=0}^{T-1}E_t^H$ and $\overline{\Delta}^H_\tau + \overline{\Delta}^A_\tau$ as $\Delta_\tau$, we obtain the following:
\begin{align}
\label{eqn:app1_eq6}
\widehat E^A_\tau &= \frac{ E^B_0 - E_{target} + \sum^{T-1}_{t=0}E^H_t -\sum^{\tau-1}_{t=0} E^A_t + \Delta_\tau}{\frac{\beta^T-\beta^\tau}{\beta^\tau (\beta-1)}} 
\end{align}
We know from Equation~\ref{eqn:EC0} that $\sum^{\tau-1}_{t=0} E^A_t = E^A_0\sum^{\tau-1}_{t=0}\beta^t$.
Since $0<\beta<1$, we have:
\begin{align}
\label{eqn:app1_eq7}
\sum^{\tau-1}_{t=0} E^A_t = \frac{\beta^\tau-1}{\beta^T-1}(E^B_0 - E_{target} + \sum^{T-1}_{t=0}E^H_t)
\end{align}
Next, we plug Equation~\ref{eqn:app1_eq7} into Equation~\ref{eqn:app1_eq6} to obtain:
\begin{align}
\label{eqn:app1_eq8}
\widehat E^A_\tau &= \frac{E^B_0 - E_{target} + \sum_{t=0}^{T-1}E_t^H}{\frac{\beta^T-1}{\beta^\tau (\beta-1)}} + \frac{ \Delta_\tau}{\frac{\beta^T-\beta^\tau}{\beta^\tau (\beta-1)}} 
\end{align}
The first term is nothing but $E^A_\tau$ as shown in Equation~\ref{eqn:app1_eq1}.
Therefore, we reach the solution provided in Lemma~\ref{lem:correction}. \qedsymbol

\section{ATUS Data and Our Pre-processing}
\label{sec:app2}
Table~\ref{tab:atusid} summarizes the activity categories in the ATUS dataset as shown in the ATUS coding lexicons document~\cite{amtus}.
We have reduced the number of categories to 10 by grouping some of the ATUS categories together and assigned them location labels as shown in Table~\ref{tab:ourid}.

\begin{table}[!h]
\caption{ATUS Category IDs and Labels}
\vspace{-2mm}
\label{tab:atusid}
\renewcommand*{\arraystretch}{0.9}
\centering
\begin{tabular}{@{}rl@{}}
\toprule
\multicolumn{2}{c}{\textbf{ATUS Major Categories}} \\ \midrule
\multicolumn{1}{c}{\textbf{ID}} & \multicolumn{1}{l}{\textbf{Label}} \\
1 & Personal care activities \\
2 & Household activities \\
3 & Caring for \& helping household members \\
4 & Caring for \& helping non-household members \\
5 & Work \& work related activities \\
6 & Education \\
7 & Consumer purchases \\
8 & Professional \& personal care services \\
9 & Household services \\
10 & Government services \& civic obligations \\
11 & Eating \& drinking \\
12 & Socializing, relaxing \& leisure \\
13 & Sports, exercise, \& recreation \\
14 & Religious \& spiritual activities \\
15 & Volunteer activities \\
16 & Telephone calls \\
18 & Travelling \\
50 & Other/invalid \\ \bottomrule
\end{tabular}
\end{table}
\begin{table}[!h]
\caption{Our assigned IDs, labels and locations.}
\label{tab:ourid}
\vspace{-2mm}
\centering
\renewcommand*{\arraystretch}{0.9}
\begin{tabular}{@{}rlll@{}}
\toprule
\multicolumn{4}{c}{\textbf{Our Categories}} \\ \midrule
\multicolumn{1}{c}{\textbf{ID}} & \multicolumn{1}{l}{\textbf{Label}} & \multicolumn{1}{l}{\textbf{ATUS IDs}} & \multicolumn{1}{l}{\textbf{Location}} \\
1 & Sleep & 1$-$1 & home \\
2 & Housework & 2 & home \\
3 & People Care & 3 \& 4 & home \\
4 & Work & 5 \& 6 & office \\
5 & Shop & 7 & store \\
6 & Eat & 11 & home/office \\
7 & Leisure & 1 (except 1$-$1) \& 12 & home \\
8 & Exercise & 13 & store \\
9 & Travel & 9 & outdoors \\
10 & Others & 8, 9, 10, 14, 15, 16 \& 50 & home \\ \bottomrule
\end{tabular}
\end{table}

\section*{Acknowledgments}
We greatly appreciate the feedback from Prof. Dimitri P. Bertsekas of Arizona State University in the preparation of this work.

\ifCLASSOPTIONcaptionsoff
  \newpage
\fi

\bibliographystyle{IEEEtran}
{\footnotesize{\bibliography{references/refs.bib}}}


\end{document}